\documentclass[pre,aps,floats,superscriptaddress,floatfix]{revtex4}
\usepackage{amssymb,amsmath}
\usepackage{amsmath,amssymb}
\usepackage{graphicx}
\usepackage{psfrag}
\usepackage{currvita}
\usepackage{siunitx}
\usepackage{xcolor}

\def\beq{\begin{equation}}
\def\eeq{\end{equation}}
\def\bea{\begin{eqnarray}}
\def\eea{\end{eqnarray}}

\begin{document}
\title{Field induced cell proliferation and death in a thick epithelium}

\author{Niladri Sarkar}\email{niladri2002in@gmail.com}
\affiliation{Max-Planck Institut f\"ur Physik Komplexer Systeme, N\"othnitzer Str. 38,
D-01187 Dresden, Germany.}\affiliation{Laboratoire Physico
Chimie Curie, UMR 168, Institut Curie, PSL Research University,
CNRS, 
Sorbonne Universiti\'e,
 75005 Paris, France.}
\author{Jacques Prost}\email{Jacques.Prost@curie.fr}
\affiliation{Laboratoire Physico
Chimie Curie, UMR 168, Institut Curie, PSL Research University,
CNRS, 
Sorbonne Universiti\'e,
 75005 Paris, France.}\affiliation{Mechanobiology Institute, National University of
Singapore, 117411 Singapore.}
\author{Frank J\"ulicher}\email{julicher@pks.mpg.de}
\affiliation{Max-Planck Institut f\"ur Physik Komplexer Systeme, N\"othnitzer Str. 38,
D-01187 Dresden, Germany.}\affiliation{Center for Systems Biology Dresden, Pfotenhauerstr. 108, 01307 Dresden, Germany}

\date{\today}
\begin{abstract}
We study the dynamics of a 
thick polar epithelium subjected to the action of both an electric and a flow field in a planar geometry.
We develop a generalized continuum hydrodynamic description and describe the tissue as a two component fluid system. 
The cells and the interstitial fluid 
are the two components and we keep all terms allowed by symmetry. 
In particular we keep track of the cell pumping activity for both solvent flow and electric current and discuss the corresponding orders of magnitude.
 We study the growth dynamics of tissue slabs, their steady states and obtain the dependence of the cell velocity, 
 net cell division rate, and cell stress on the flow strength and the applied electric field. 
 We find that finite thickness tissue slabs exist only in a restricted region of phase space and that relatively modest electric fields or imposed external flows can induce either proliferation or death.
\end{abstract}

\maketitle

\section{Introduction}
\label{intro}

During the development of an organism from a fertilized egg, tissues are formed by the collective organization of many cells
that divide or undergo apoptosis. Tissues grow by repeated rounds 
of cell division \cite{wolpert1998}. Cell apoptosis or programmed cell death plays a vital role in maintaining tissue homeostasis, and 
suppression of apoptosis can result in abnormal cell proliferation which might turn into cancer \cite{weinberg2013}. 
The processes of cell division and apoptosis occur during all stages of development and cells removed by 
apoptosis 
are often continuously replaced by cell division. Many studies aim to understand how gene regulatory pathways 
and biochemical signalling are involved in the regulation and coordination of division and apoptosis 
\cite{alberts2007,kinchen2010,buchakjian2010}. In recent times 
the mechanical properties of epithelia 
have been a topic of great interest, and studies have shown how tissue growth is regulated by local pressure 
or stiffness \cite{shraiman2005,mammoto2010}, and how cell division and apoptosis depend on it 
\cite{montel2011,montel2012,delarue2013}. The stress at which cell division balances cell apoptosis on 
average is called the homeostatic 
stress of the tissue and in that state the tissue can remain 
stationary \cite{basan2009}. Perturbations from the homeostatic state lead to interesting tissue dynamics that can affect tissue morphogenesis. The concept of stress dependent cell division and apoptosis plays an important role in the present work. 

To better understand the mechanical properties of tissues, different theoretical approaches have been
developed ranging from 
individual cell based models \cite{montel2011,drasdo2005,podewitz2016,alt2017} to hydrodynamic continuum descriptions \cite{araujo2004,bittig2008,byrne2009,blanch2014,risler2015,hannezo2016,popovic2017,komura2018,savin2018}. 
Coarse grained hydrodynamic models can be used to understand the multicellular dynamics of tissues at long wavelengths. In \cite{ranft2010}, a one component continuum description of tissues has been 
developed which takes into account the stress distribution and the flow field generated by cell division 
and apoptosis and shows that the tissue effectively behaves as a viscoelastic fluid at long time scales. 
In \cite{ranft2012}, a two component fluid description of tissues is developed, which considers the cells 
together with the extracellular matrix as one component and the interstitial fluid as the other. 
It takes 
into account the material turnover as a result of cell division and apoptosis explicitly, but also 
the permeation of the interstitial fluid through a tissue. 

Fluid transport in a tissue plays an important role in the mechanical signalling during 
organogenesis 
\cite{gilbert2016}. If the fluid flow mechanics is altered, gene expression in cells is affected, 
leading to disruption in organ development and congenital malfunctions. The cells in a tissue are active and 
can pump fluid through it. Direct in-vitro experiments on 
rabbit corneal epithelial tissue have shown that cells can pump fluid against a huge 
pressure difference \cite{maurice1972}. Insect malpighian tubules have been seen to pump fluids and ions dependent on the activity of $H^+$ transporting V-ATPase which resides on the lumen \cite{maddrell1992}. Studies on corneal endothelium have shown that 
lactate flux is responsible for pumping fluids through them \cite{li2016}. Though a significant amount of experimental evidence is present in the literature, theoretical studies involving fluid pumping mechanism in an epithelium is lacking. In this article, we have developed a hydrodynamic theory of epithelium in which the cells can pump fluid through the tissue 
due to active processes such as ion pumping.

 Specific pumps can generate flows of ions of both positive and negative charges inside 
  a tissue. 
  The corresponding electric current gives rise to an internal electric field 
  which generates a potential drop across the tissue. Several 
studies have shown the presence of electrical voltage difference across tissues 
\cite{cereijido1978,josephson1979,hay1985,hamilton2016}. 
Here we consider growth dynamics and steady state of a layer of cells, which represents a thick epithelium, in the presence 
of electric currents and permeating flows. We work at constant  electric current which can be imposed by a corresponding external 
electric potential difference. Similarly we work at imposed flow velocity. The flow velocity can be imposed by an external
pressure difference, see Fig. 1.

This paper is organized as follows. Section II introduces the different conserved quantitites relevant for a 
thick epithelium. The constitutive equations describing tissue material properties and tissue dynamics based on symmetry 
considerations are introduced in Section III. A model for a thick permeating epithelium is discussed as an example 
in Section IV, where we  analyse the dynamics and steady state properties. We conclude our findings in Section V.  

\section{Conservation laws}

We first present the hydrodynamic theory of a polar tissue that is permeated by
interstitial fluid flows in the presence of an electric field, starting with a discussion of conservation laws.

\subsection{Volume Conservation}

In our coarse grained approach, we define the cell volume as $\Omega^c$, and the fluid molecular volume as 
$\Omega^f$. The number of cells per unit volume is denoted as $n^c$, and the number of interstitial fluid 
molecules per unit volume is denoted as $n^f$. If the number of cells in a volume $V$ is $N^c$, and the 
number of interstitial fluid molecules $N^f$, then $n^c=N^c/V$, and $n^f=N^f/V$, and we have 
$V=N^c\Omega^c + N^f\Omega^f$ \cite{ranft2012} or equivalently,
\bea
n^c\Omega^c + n^f\Omega^f=1. \label{vol}
\eea
The cell volume fraction is defined as $\phi=n^c\Omega^c$. The volume fraction of the interstitial  fluid is then given by $\phi^f=1-\phi=n^f\Omega^f$.  

The cell number density $n^c$ obeys a balance equation which has a flux contribution from cell flows, and also 
a source/sink contribution from 
cell division and apoptosis \cite{ranft2010},
\bea
\partial_tn^c + \partial_\alpha(n^c v^c_\alpha)=n^c(k_d-k_a), 
\label{cellcon}
\eea
where $v^c_\alpha$ is the cell velocity, and, $k_d$ and $k_a$ are the rates of cell division and apoptosis respectively. In a hydrodynamic description, we allow the cells to be converted to interstitial fluid when they undergo apoptosis and vice versa when they grow and divide. The balance equation for the interstitial fluid can then be written as
\bea
\partial_tn^f +\partial_\alpha(n^f v^f_\alpha)=-{\Omega_c \over \Omega_f}n^c(k_d-k_a) - n_c{d \over dt} {\Omega_c \over \Omega_f}, \label{flucon} 
\eea
where $v^f_\alpha$ is the velocity of the interstitial fluid, and, ${\Omega_c \over \Omega_f}n^c$ is the volume of fluid particles corresponding to the volume of one cell assuming for simplicity that 
the fluid and cell mass densities are the same \cite{ranft2012}. 

\subsection{Charge Conservation}

The conservation of charges can be written as
\bea
{\partial \rho \over \partial t}+\nabla\cdot{\bf I}=0, \label{cont}
\eea
where $\rho({\bf x},t)$ is the local charge density, and ${\bf I}({\bf x},t)$ is the electric current. 

\subsection{Momentum conservation}


We consider the total stress to be given $\sigma_{\alpha\beta}=\sigma^c_{\alpha\beta}
+ \sigma^f_{\alpha\beta}$, where $\sigma^c_{\alpha\beta}$ is the stress associated with the cells and $\sigma^f_{\alpha\beta}$ is the stress associated with the interstitial fluid. The cell 
stress can be decomposed into an isotropic part $\sigma^c$ and a traceless anisotropic part 
$\tilde\sigma^c_{\alpha\beta}$, so that the total cell stress 
can be written as
\bea
\sigma^c_{\alpha\beta}=\tilde\sigma^c_{\alpha\beta} + \sigma^c\delta_{\alpha\beta}.
\label{cellstress}
\eea
For simplicity, we consider the anisotropic stress in the interstitial fluid to vanish 
over length scales large compared to that of the cells. Then the fluid stress is of the form 
\bea
\sigma^f_{\alpha\beta}=-P^f\delta_{\alpha\beta}.
\label{fluidstress}
\eea

The force balance 
of the tissue in the absence of any external forces captures momentum conservation and can be written as  
\bea
\partial_\beta(\sigma^c_{\alpha\beta}+\sigma^f_{\alpha\beta})=0.
\label{forcebal}
\eea
The force balance in Eq. (\ref{forcebal}), allows for a momentum transfer between the cells and the 
interstitial fluid which corresponds to an internal force \cite{ranft2012}. 
As a consequence we write 
\bea
\partial_\beta\sigma^c_{\alpha\beta} + f_\alpha &=& 0, \label{cellforcebal} \\
\partial_\beta\sigma^f_{\alpha\beta} - f_\alpha &=& 0. \label{fluidforcebal}
\eea
Here $f_\alpha$ is the force which takes into account the momentum exchance between the two components of the system. 

\section{Constitutive equations}

We now introduce the constitutive equations of a polar tissue permeated by a fluid and subject to an electric 
field. We consider a tissue in which the cells are polar, i.e., they exhibit a structural 
polarity that can be characterised by a unit 
 vector ${\bf p}$ with $p^2=1$. This further implies the existence of a nematic order parameter 
for the 
tissue given by $q_{\alpha\beta}=p_\alpha p_\beta - (1/3)p^2\delta_{\alpha\beta}$. Furthermore the 
cells can generate active stress as a result of active processes in the cytoskeleton, such as the action of molecular motors which 
consume a chemical fuel ATP. This generated stress can additionaly have a contribution from cell division and apoptosis 
 \cite{ranft2010,Delarue2014}. 

To start with we write the constitutive equations for the cell volume and the cell volume fraction to be 
\bea
\Omega^c=\Omega^c(\sigma^c,q_{\alpha\beta}\tilde\sigma^c_{\alpha\beta},p_\alpha E_\alpha), \,\,\,\mbox{and} \,\,\, \phi=\phi(\sigma^c,q_{\alpha\beta}\tilde\sigma^c_{\alpha\beta},p_\alpha E_\alpha). \label{eos}
\eea 
where $\Omega_c$ and $\phi$ are functions of the isotropic cell stress, $\sigma^c$, the projection of 
the anisotropic cell stress $\tilde\sigma^c_{\alpha\beta}$ on the nematic order parameter $q_{\alpha\beta}$, and the 
projection of the applied electric field $E_\alpha$ on the cell polarity $p_\alpha$. 
The osmotic compressibility of the cells is defined as $\chi^{-1}=(n^c)^{-1}(dn^c/d\sigma^c)$. Because
$n^c={\phi/\Omega^c}$, we also have $\chi=n^c[d(\phi/\Omega_c)/d\sigma^c]^{-1}$. The expansion of 
$dn^c/dt$ to first order using eq. (\ref{eos}) can be written as
\bea
{1 \over n^c}{dn^c \over dt}= -\chi^{-1}{d\sigma^c \over dt} - \chi_2^{-1} {d(q_{\alpha\beta}\tilde\sigma^c_{\alpha\beta}) 
\over dt} -\chi_3^{-1} {d(p_\alpha E_\alpha) \over dt} . \label{sig1} 
\eea 
where $\chi_2=n^c[d(\phi/\Omega_c)/d(q_{\alpha\beta}\tilde\sigma^c_{\alpha\beta})]^{-1}$, and 
$\chi_3=n^c[d(\phi/\Omega_c)/d(p_\alpha E_\alpha)]^{-1}$.
Eq.~(\ref{cellcon}) can then be rewritten as
\bea
{1 \over n_c}{dn_c \over dt}=-v^c_{\gamma\gamma}+k_d - k_a \; , 
\label{isostressrate}
\eea

The net growth rate $k_d-k_a$ of the tissue is in general a function of 
$\sigma^c$, $q_{\alpha\beta}\tilde\sigma^c_{\alpha\beta}$, $p_\alpha E_\alpha$,
where we focus our attention on variables that are even under time reversal.
It can be expressed to first order 
as
 \cite{ranft2012} 
\bea
k_d-k_a=\bar\eta^{-1}(P^c_h + \sigma^c + \nu \tilde\sigma^c_{\alpha\beta}
q_{\alpha\beta} + \nu_1 p_\alpha E_\alpha), \label{cellgrowthrate}
\eea
where $\bar\eta^{-1}=d(k_d-k_a)/d\sigma^c|_{\tilde\sigma^c_{\alpha\beta},E}$, $\nu$ and $\nu_1$ are expansion coefficients.  The homeostatic pressure of the tissue in the absence of anisotropic stress and electric field is denoted $P^c_h$. Using equations (\ref{isostressrate}) and (\ref{cellgrowthrate}), we obtain 
a general constitutive equation for the isotropic cell stress
\bea
\left(1+\tau_l{d \over dt}\right)(\sigma^c + P^c_h) + \nu \left(1+\tau_\nu {d \over dt}\right)\tilde\sigma^c_{\alpha\beta}q_{\alpha\beta} + \nu_1 \left(1+\tau_1{d \over dt}\right)p_\alpha E_\alpha= \bar\eta 
v^c_{\gamma\gamma}, \label{isostressform1}
\eea
where $\tau_l=\bar\eta/\chi$ is the isotropic relaxation rate, $\tau_\nu=\bar\eta/\chi_2$ is the anisotropic relaxation rate, $\tau_1=\bar\eta/\chi_3$ is the relaxation rate arising from the electric field, and $\bar\eta$  is 
the bulk viscosity. For slowly varying states
we can neglect relaxation processes and Eq. (\ref{isostressform1}) simplifies to
\bea
\sigma^c + P^c_h = \bar\eta 
v^c_{\gamma\gamma} - \nu \tilde\sigma^c_{\alpha\beta}q_{\alpha\beta} - \nu_1 p_\alpha E_\alpha. \label{isostressform}
\eea
 
Similarly we can write on symmetry grounds a constitutive equation for the anisotropic part of the cell stress which reads to linear order
\bea
\tilde\sigma^c_{\alpha\beta}=2\eta \tilde v^c_{\alpha\beta}+\zeta q_{\alpha\beta} - \nu_2(p_\alpha E_\beta + p_\beta E_\alpha - {2 \over 3}p_\gamma E_\gamma \delta_{\alpha\beta}), \label{anisocellstressform}
\eea
where for simplicity we introduce an isotropic shear viscosity $\eta$ and $\nu_2$ is a coefficient, which couples the 
electric field to the anisotropic cell stress.  The active anisotropic part of the  cell stress is given by 
$\zeta q_{\alpha\beta}$. Similarly, we write a constitutive equation for the momentum exchange $f_\alpha$ 
between interstitial fluid and cells 
\bea
f_\alpha=-\kappa(v^c_\alpha - v^f_\alpha) + \lambda_1 p_\alpha + \lambda_2 E_\alpha + \lambda_3q_{\alpha\beta}E_\beta +
\lambda_4 \partial_\beta q_{\alpha\beta},
\label{falpha}
\eea
where $\kappa$ describes friction between the cells and the fluid and $\kappa^{-1}$ is the effective permeability of the tissue. The  term caracterized by the coefficient $\lambda_1$ represents the force density exerted by the interstitial fluid 
on the cells as a result of the pumping activity of  the cells. 
 The third and fourth terms represent the isotropic and anisotropic parts of the force density generated by electric fields. They are characterized by  the  coefficients $\lambda_2$ and $\lambda_3$  respectively. 
   The fifth term characterized by the coefficient $\lambda_4$ represents the contribution to $f_\alpha$ arising from the gradient of the nematic order parameter. 
Finally we can also write a constitutive equation for the electric current 
\bea
I_\alpha=-\bar\kappa(v^c_\alpha - v^f_\alpha) + \Lambda_1 p_\alpha + \Lambda_2 E_\alpha + \Lambda_3q_{\alpha\beta}E_\beta +
\Lambda_4 \partial_\beta q_{\alpha\beta}, 
\label{ialpha}
\eea   
where $\bar\kappa$ 
is the coefficient describing the streaming current, $\Lambda_1$  the coefficient 
describing the current resulting from a polar distribution of ion pumps, and $\Lambda_2$, 
 $\Lambda_3$ are respectively the isotropic and anisotropic part of the electric conductivity tensor. 
  The coefficient $\Lambda_4$ is an out-of-equilibrium flexoelectric coefficient. 
On the coarse-graining scale we use here, the tissue is neutral. The local electric equilibrium is
fast 
 compared to tissue dynamics. Thus $\partial_t \rho\simeq 0$, which
imposes a condition of conservation of the electric current (\ref{ialpha}) 
\bea
\partial_\alpha I_\alpha=0. \label{divi}
\eea

Furthermore, assuming the cell and fluid mass densities to be equal and constant  \cite{ranft2012} 
 implies that the total volume flux $v_\alpha=n^c\Omega^cv^c_\alpha + n^f\Omega^fv^f_\alpha$ is divergence free 
\bea
\partial_\alpha v_\alpha=0.  \label{incomp}
\eea 
The constraint of incompressibility (\ref{incomp}) is imposed by using the fluid pressure $P^f$ as a Lagrange multiplier. In a similar way, the constraint of current conservation (\ref{divi}) is imposed by using the electric potential $U$, such that $E_\alpha=-\partial_\alpha U$,
as the Lagrange multiplier.

\section{Thick epithelium on a permeable substrate}
\label{full}

\begin{figure}[htb]
\includegraphics[height=7cm]{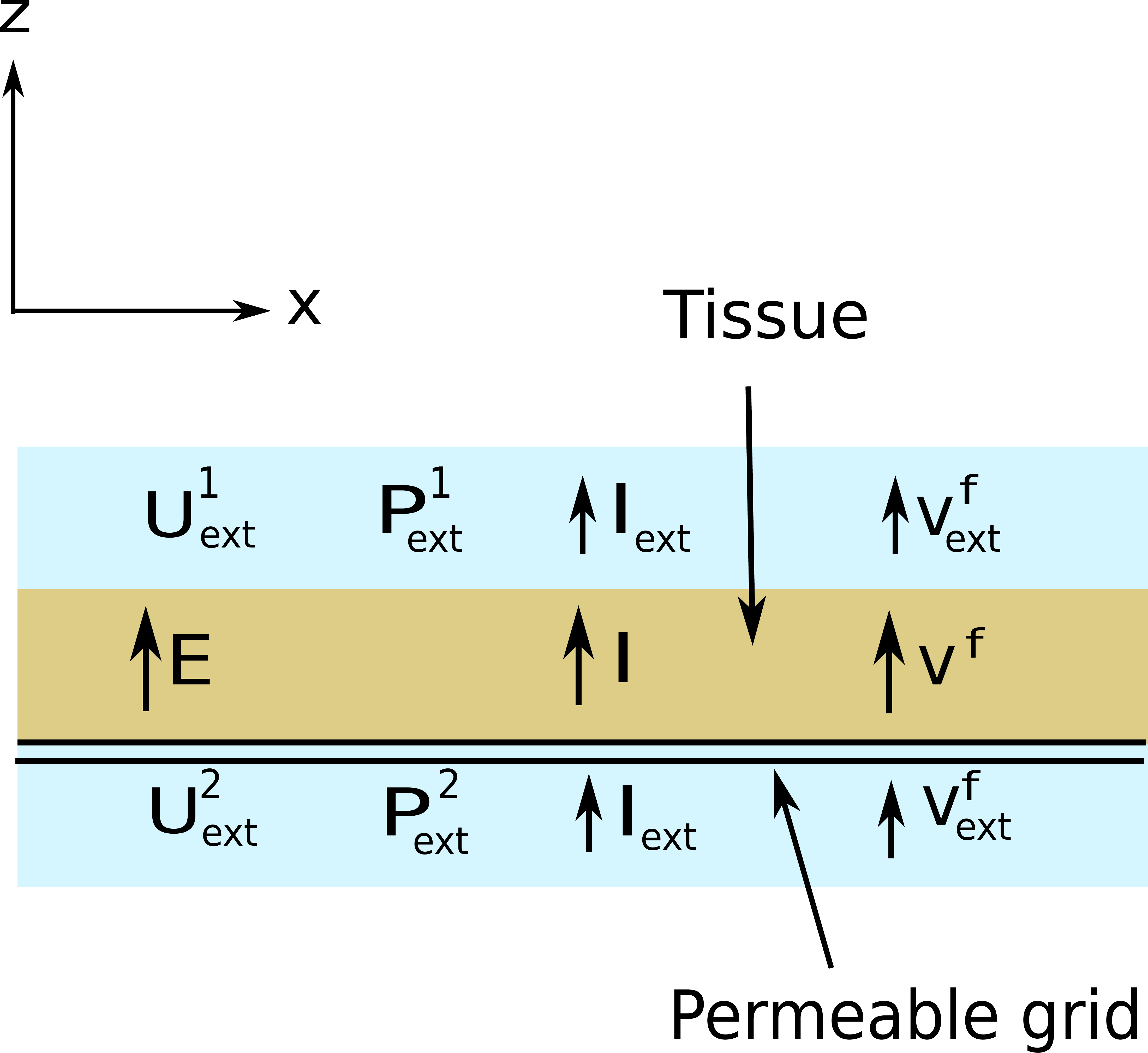}
\caption{
 Schematic diagram of a thick 
epithelium placed on a porous substrate. 
The tissue layer of height $h$ is shown in brown, the two external fluid components on both sides of the 
tissue are shown in blue. The double black line features the
permeable grid on which the tissue is placed. It is permeable to fluid and ions but impermeable to cells. The 
electric potential at both sides of the tissue are denoted by $U^{1,2}_{\rm ext}$. The electric field across the tissue is denoted by $E$ and the electric current by $I$. The difference of the external hydrostatic pressure 
$P^{1,2}_{\rm ext}$ corresponds to a normal force per unit area that acts on the grid. The tissue is permeated by a fluid at a velocity $v^f$. 
 }
\label{fullmodel}
\end{figure}

We consider a thick planar tissue consisting of cells and interstitial fluid resting on a substrate and embedded in a fluid 
medium. 
The fluid surrounds the tissue-substrate system from all sides and can permeate the substrate whereas cells cannot. 
A constant fluid flow $v^f_{\rm ext}$ through  the tissue is imposed for instance by an appropriate hydrostatic
pressure difference. 
Similarly, we work at constant imposed electric current density $I_{\rm ext}$.
We study how the interplay of cell division, cell apotosis, fluid pumping, osmotic pressure and electric current in the tissue controls 
its dynamics and morphology. 
We only consider a homogeneous slab, so that dynamical quantities depend only on one variable which we call z
and which describes the distance from the substrate. 
We postpone for future work the study of the stability of the found solutions with respect to variations of quantities parallel to the slab. 
We consider cells uniformly polarised along the z-direction,  and we chose $p_z=1$. A schematic
representation of the system is given in Fig.~\ref{fullmodel}

\subsection{Constitutive equations in planar geometry}

We first write down the constitutive equations for our model tissue in one dimension. 
Using  (\ref{isostressform}) and  (\ref{anisocellstressform}), the constitutive equations for the cell stress $\sigma^c_{zz} $ can be written as
\bea
\sigma^c_{zz} = -P^H_0 + \eta_{\mathrm{eff}}\partial_z v^c_z - \nu_{\mathrm{eff}}E_z, \label{totcellstr}
\eea
where $P^H_0=P^c_h - 2\zeta(1-2\nu/3)/3$ is the effective homeostatic pressure of the tissue, defined as the difference between the homeostatic pressure and the total active pressure of the tissue, and $\eta_{\mathrm{eff}}=\bar\eta + 4\eta(1-2\nu/3)/3$ is the effective viscosity of the tissue. Both of them are constants for the given system. 
Here $\nu_{\mathrm{eff}}=\nu_1 + 4\nu_2(1-2\nu/3)/3$ is the effective 
 coefficient coupling electric field and stress. 
Using Eqs. (\ref{fluidstress}), (\ref{fluidforcebal}), and (\ref{falpha}), the force balance equation 
 takes the form
\bea
\partial_zP^f_z=\kappa (v^c_z-v^f_z) -\lambda_1 -(\lambda_2 +{2 \over 3}\lambda_3) E_z. \label{fluforcebaltot} 
\eea
Using  
the current conservation law (\ref{divi}) we have $I_z(z)=I_{\mathrm{ext}}$, 
where $I_{\mathrm{ext}}$ the externally imposed electric current and 
the tissue electric field reads
\bea
E_z &=& {I_{\mathrm{ext}}-\Lambda_1 \over \Lambda} + {\bar\kappa (v^c_z -v^f_z) \over \Lambda}. 
\label{elec}
\eea   
where $\Lambda=\Lambda_2 + 2\Lambda_3/3$ is the electric conductivity.

\subsection{Boundary conditions}
\label{bcs}

We specify boundary conditions for a planar epithelium of thickness $h$. 
We allow for narrow surface layers of thickness $e$, comparable to cell size,
 at both $z=0$ and $z=h$ to have have growth rates $k_d-k_a$ that differ from the bulk values by a value $\delta k$. This gives rise to a difference between cell velocity
  and interface velocity near the interface.  Thus the cell velocity at $z=h$ is related
 to the  interface velocity by
\bea
v^c_z(z=h)={dh \over dt} - v_{1} \label{velcbc1}
\eea
where $v_1=\delta k_1 e$. At $z=0$ the tissue is attached to a immobile substrate. The
boundary condition for the cell velocity  reads
\bea
v^c_z(z=0)=v_2. \label{velcbc2}
\eea
with $v_2=\delta k_2 e$. 

Incompressibility of the tissue (\ref{incomp}) implies $\phi v^c_z + (1-\phi)v^f_z=v^f_{\mathrm{ext}}$, where $v^f_{\mathrm{ext}}$ is the externally imposed
 velocity of the external fluid. Therefore the interstitial fluid velocity $v_z^f$ reads
\bea
v^f_z = {v^f_{ext} \over 1-\phi} - {\phi \over 1-\phi}v^c_z. 
\label{fluidvel}
\eea
Force balance at the tissue surface implies
stress continuity.  Thus the total stress is balanced by the hydrostatic pressure $P_{\mathrm{ext}}^1$
of the external fluid at $z=h$ 
\bea
\sigma_{zz}(z=h)=\sigma^c_{zz} - P^f=-P_{\mathrm{ext}}^1, \label{strbaltot} 
\eea

At $z=h$ fluid is exchanged with the tissue at a rate that is driven by
the fluid chemical potential difference between the upper outside region (denoted with the index 1, see Fig. 1) and the tissue, 
and can be written as
\bea
\Lambda^f[(P_{\mathrm{ext}}^{1}-P^f)-\Pi_{\mathrm{ext}}^{1}]
+J_p={dh \over dt}-v^f_{\rm ext} , 
\label{chembal}
\eea 
Here, $\Pi_{\mathrm{ext}}^{1}$ denotes the external osmotic pressureb of osmolites
that do not to enter the tissue or the interstitial fluid and $\Lambda^f$ describes the
permeability of the interface for water flow. 
This is for example the case for dextran with molecular mass exceeding $100 kDa$ \cite{montel2011}. 
The flux $J_p=\Lambda^f
(\Pi_{\mathrm{ext},0}^{1}-\Pi_{\mathrm{int},0}^{1})$, can be nonzero as a result of active pumps and transporters. It looks like
a flow due to an effective water pump. Here $\Pi_{\rm ext,0}^1$ and
$\Pi_{\rm int,0}^1$ denote
the outside and inside osmotic pressure, respectively, 
of osmolites that can be exchanged  between external fluid
and tissue.  
Using Eq. (\ref{chembal}), the total cell stress  at $(z=h)$ reads
\bea
\sigma^c_{zz}=-\Pi_{\mathrm{ext}}^1+\frac{J_p}{\Lambda^f} +
{1 \over \Lambda^f}\left(v^f_{ext}-{dh \over dt} \right)
. \label{cellstresstop}
\eea
Using the above boundary conditions we now discuss the thickness dynamics of
a thick epithelium on a substrate or basal membrane.

\subsection{Internal dynamics of a thick epithelium}

\subsubsection{Time dependence of tissue thickness}
\label{noelec}

We first write the equation for the cell velocity $v^c_z$ for the case of constant
cell volume fraction $\phi$:
\bea
\lambda_0^2\partial_z^2v^c_z - a\partial_zv^c_z -v^c_z + v_{\Lambda}=0, \label{forcebalfinal} 
\eea
where $\lambda_0$ with  $\lambda_0^2=\eta_{\mathrm{eff}}(1-\phi) /\kappa_{\mathrm{eff}}$  is a hydrodynamic screening length discussed in ref. \cite{ranft2012}. The length  $a={\nu_{\mathrm{eff}}\bar\kappa / (\Lambda
\kappa_{\mathrm{eff}})}$  stems from the influence of the electric field on tissue flow. The effective fluid pumping velocity is
given by $v_\Lambda={[{\lambda
(I_{\mathrm{ext}} -\Lambda_1) / \Lambda} +\lambda_1](1-\phi) / \kappa_{\mathrm{eff}}} +v^f_{\mathrm{ext}}$ and $\kappa_{\mathrm{eff}}=\kappa- {\lambda\bar\kappa/ \Lambda}$ is an effective permeability where $\lambda=\lambda_2 + 2\lambda_3/3$. 

Equation (\ref{forcebalfinal})  can be solved for given boundary conditions (\ref{cellstresstop}) and (\ref{velcbc2}), to determine the velocity profile and the
coresponding cell stress profile 
\bea
v^c_z(z) &=& v_\Lambda + (v_2 - v_\Lambda)\exp(k_1z) \nonumber \\
&+&  (\exp(k_2z)-\exp(k_1z)){
\bar P^H/\eta_{\mathrm{eff}} + a v_\Lambda/\lambda_0^2 + (a/\lambda_0^2 -k_1)(v_2 - v_\Lambda)
\exp(k_1h) \over (k_2-a/\lambda_0^2)\exp(k_2h)- (k_1-a/\lambda_0^2) \exp(k_1h)},  
\nonumber \\ \label{vcfull} \\
\sigma^c_{zz} &=& - \sigma^0_{zz} + \eta_{\mathrm{eff}}\left[\left(k_1-{av_\Lambda \over \lambda_0^2}\right)
(v_2 - v_\Lambda)
\exp(k_1z) - {av_\Lambda \over \lambda_0^2}\right]
 \nonumber \\
 &+&  {[\bar P^H + \eta_{\mathrm{eff}}av_\Lambda/\lambda_0^2 - \eta_{\mathrm{eff}}(k_1- a/\lambda_0^2)
 (v_2 - v_\Lambda)\exp(k_1h)] \over [(k_2- a/\lambda_0^2)\exp(k_2h) - (k_1-a/\lambda_0^2)\exp(k_1h)]} 
 \nonumber \\
&& \times[(k_2- a/\lambda_0^2)\exp(k_2z) - (k_1-a/\lambda_0^2)\exp(k_1z)], \label{strcfull} 
\eea
where $k_{1,2}={(a\pm\sqrt{a^2+4\lambda_0^2}) / 2\lambda_0^2}$. 
Here $\bar P^H=P^c_h-2\zeta(1-2\nu/3)/3+\nu_{\rm eff}
[(I_{\mathrm{\rm ext}}-\Lambda_1)/\Lambda - 
\bar\kappa v^f_{\mathrm{\rm ext}}\nu_{\rm eff}/[\Lambda(1-\phi)]]- \Pi^1_{\rm ext}
+ (J_p+v^f_{\rm ext})/\Lambda^f$ is an effective homeostatic pressure of the tissue which  takes into account the
effects of electric currents and fluid flows. Eqs. (\ref{vcfull}) and (\ref{strcfull}) show that the instantaneous cell velocity and the cell stress profiles combine two exponentials with
characteristic lengths $k_{1,2}^{-1}$.  

Using the boundary condition (\ref{velcbc1}) for the cell velocity at  $z=h$, we obtain a dynamical equation for the tissue thickness:
\bea
{dh \over dt} &=& {v_\Lambda \over (1+\Lambda_h)}  - {[\exp{(k_2h)}- \exp{(k_1h)}] \over (1+\Lambda_h)}
{[(k_1 - a/\lambda_0^2)
(v_2-v_\Lambda)\exp(k_1h) - av_\Lambda/\lambda_0^2  - \bar P^H/\eta_{\mathrm{eff}}] \over 
[(k_2-a/\lambda_0^2)\exp{(k_2h)} - (k_1-a/\lambda_0^2)\exp{(k_1h)}]} \nonumber \\ 
&& + {(v_2-v_\Lambda)\exp(k_1h) \over (1+\Lambda_h)}, \label{dhdt1}
\eea
where $\Lambda_h=[\exp(k_2 h)-\exp(k_1 h)]/[\Lambda^f\eta_{\mathrm{eff}}((k_2-a/\lambda_0^2)\exp(k_2h)
- (k_1-a/\lambda_0^2)\exp(k_1h))]$ is non-zero and positive.

Eq. (\ref{dhdt1}) can give rise to three different behaviours, tissue growth, tissue collapse or finite steady state thickness. 
If on one hand cell division dominates over cell apoptosis $dh/dt$ is positive and tissue growth takes place. 
If $dh/dt$ remains positive at all times, 
the thickness increases indefinitely, leading to a 
complete invasion of the available space. This corresponds to what is commonly called
tissue proliferation.
If on the other hand apoptosis dominates over cell division, then $dh/dt$ is negative, the tissue shrinks. 
If $dh/dt<0$ persists at all times, the tissue finally collapses. The tissue can also reach a stable steady state with $dh/dt=0$ and
constant thickness. The steady state thickness then depends on the 
imposed electric current and fluid flow. Steady states can also be unstable. In this case the tissue thickness 
 evolves away from the steady state value. This can give rise to either tissue collapse, indefinite growth or finite 
 thickness, depending on initial conditions and parameter values.

For simplicity, we restrict our analysis to the case $a\ll \lambda_0$. This
is motivated by simple estimates of $\bar\kappa\simeq  10^3$C/m$^3$, $\nu_{\rm eff}\simeq 3\cdot 10^{-2}$Pa m/V, and $\Lambda\simeq
6\cdot 10^{-3}\Omega^{-1}$m$^{-1}$ for which $a\simeq 5 \cdot 10^{-10}$m. We also estimate 
 $\lambda_0\simeq 1$cm, see appendix. 
For $a\ll \lambda_0$, we have $k_1=-k_2=1/\lambda_0$.  
 In this case the time dependence of tissue thickness simplifies to
 \bea
 {(1+\Lambda_h) \over v_\Lambda}\frac{dh}{dt} &= f(h)=&1 - {\alpha \over \cosh(h/\lambda_0)}+\beta 
 \tanh(h/\lambda_0), 
 \label{dhdt2}
 \eea
 where $\alpha=1- v_2/v_\Lambda$, and $\beta=\bar P^H\lambda_0/(\eta_{\rm{eff}}v_\Lambda)$.


\begin{figure}[htb]
\includegraphics[width=6cm,height=4cm]{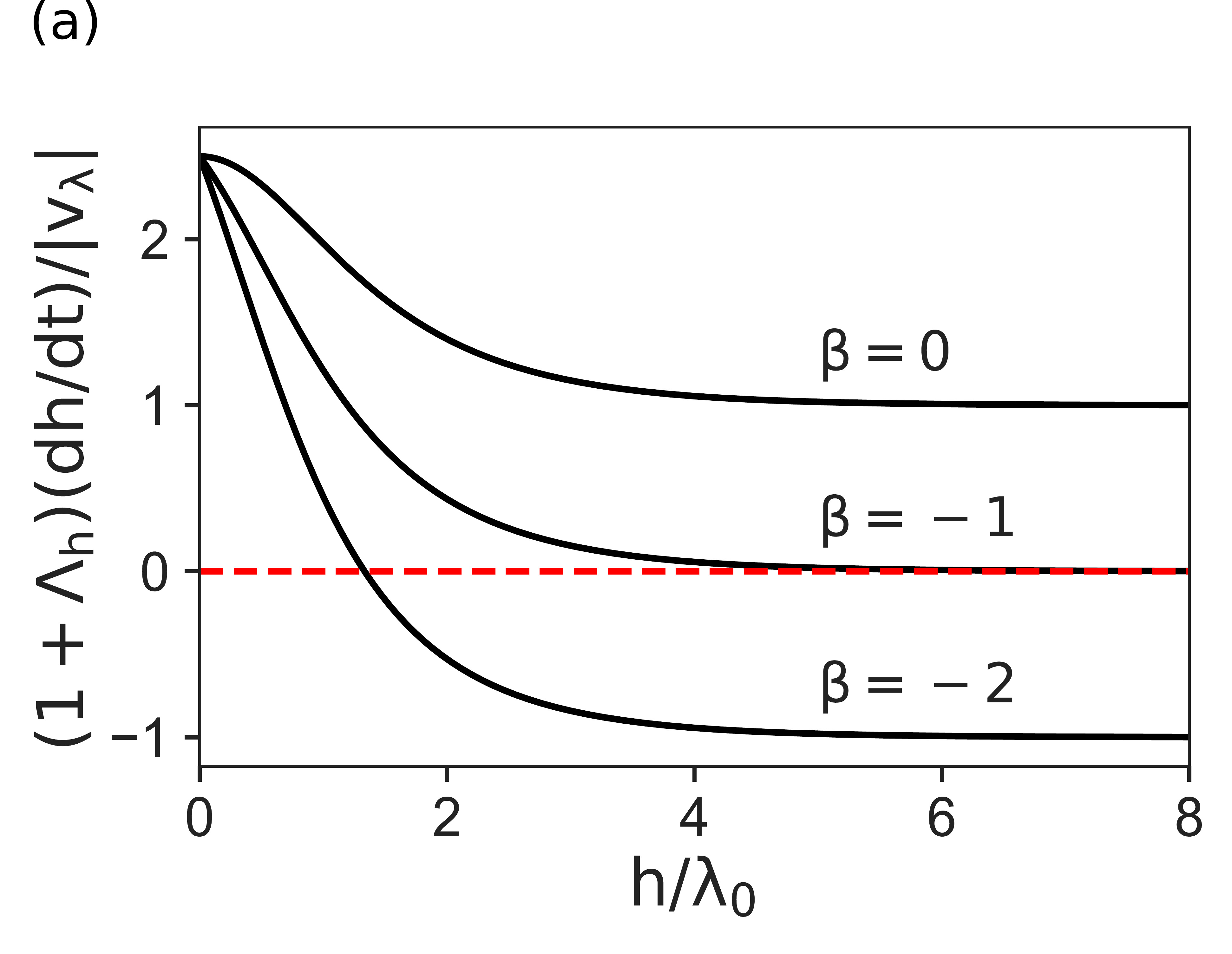}
\includegraphics[width=6cm,height=4cm]{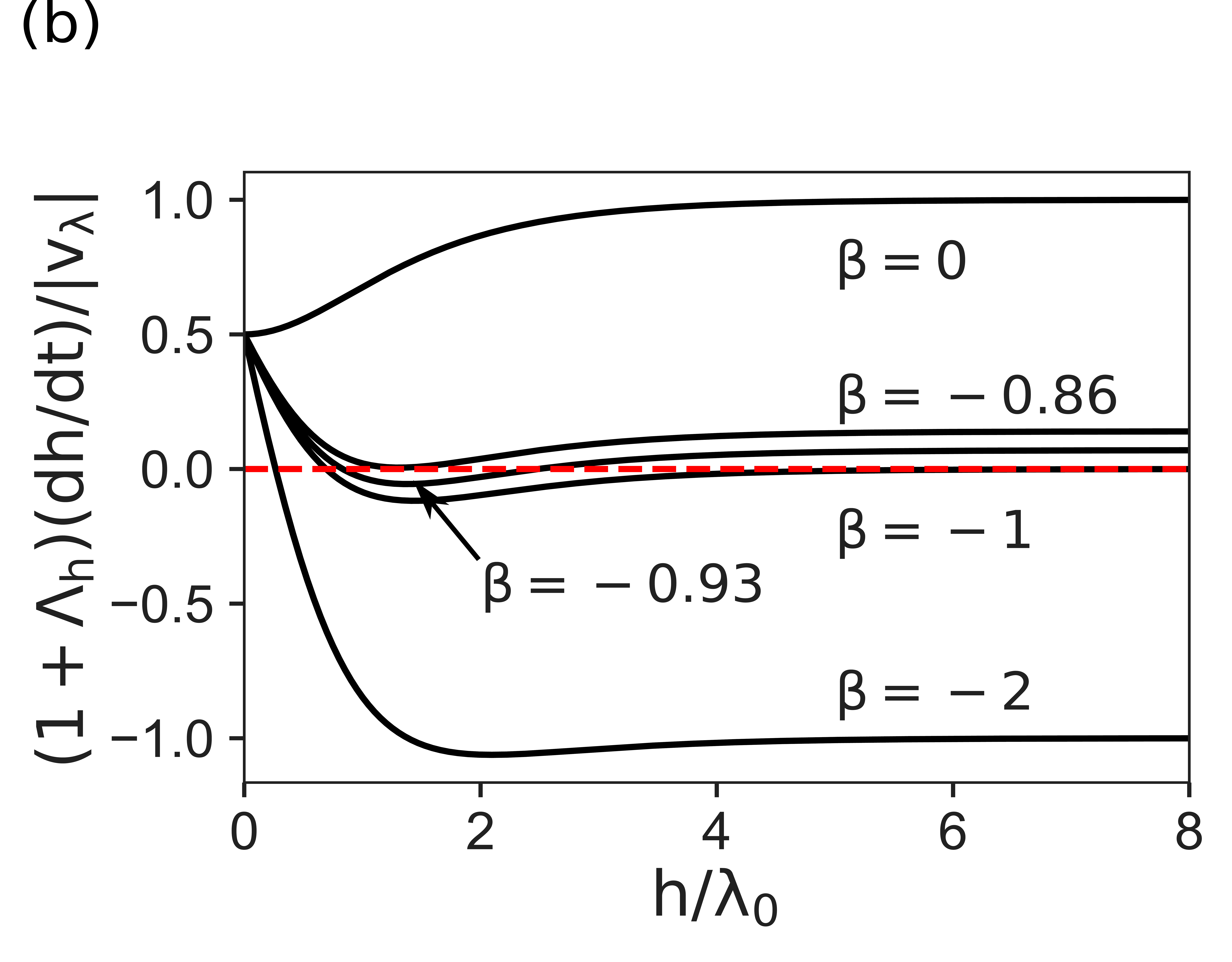}
\includegraphics[width=6cm,height=4cm]{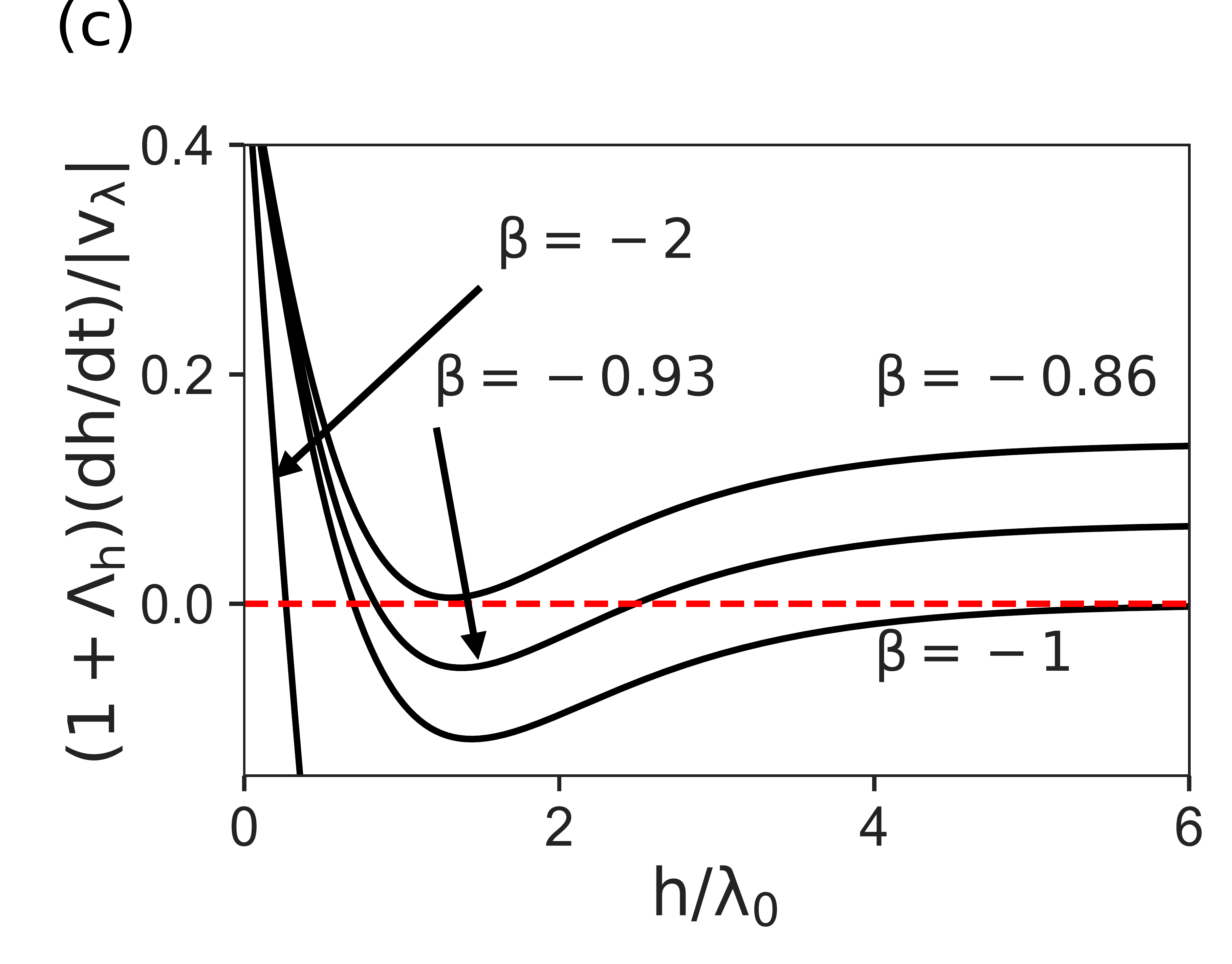}
\includegraphics[width=6cm,height=4cm]{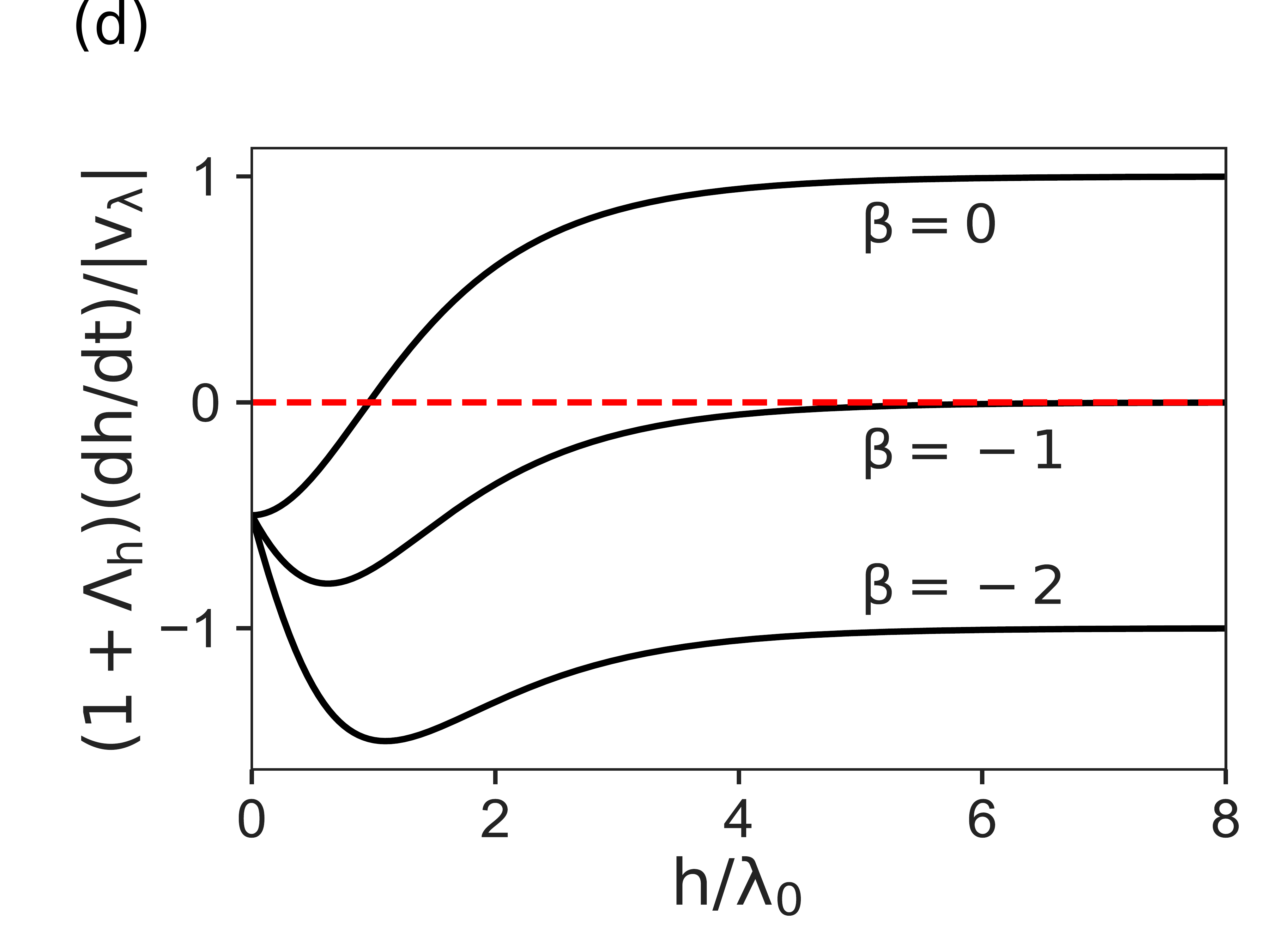}
\caption{Plot of the ratio ${(1+\Lambda_h) \over \mid v_\Lambda \mid}\frac{dh}{dt}$ as a function of $h$ for different values of $\alpha$ and $\beta$.  For positive  $v_\Lambda$, a positive (negative) value of this ratio corresponds to a
 thickness increase (decrease) of the epithelium layer; 
 for negative $v_\Lambda$ a positive (negative) value of this ratio corresponds to a decrease 
 (increase) of the layer thickness. 
(a) 
the three curves corresponding to $\beta=-2,-1$ and $0$ are generic for all $\alpha<0$. For all $\beta<-1$ the curve intersects the $h$ axis once for a finite $h=h_{ss}$ value, leading to a stable steady state if $v_\Lambda>0$, unstable steady state if $v_\Lambda<0$. As $\beta \to -1$, the steady state thickness diverges 
$h_{ss} \to \infty$, wich signals a continuous transition taking place for $\beta=-1$. For $\beta>-1$ and 
$v_\Lambda>0$ proliferation takes place and for $\beta>-1$ and $v_\Lambda<0$ collapse is predicted.
(b and c) the five curves corresponding to  $\beta=-2,-1,-0.93,-0.86$ and $0$ are typical of the succession of curve shapes for $0<\alpha<1$; the intersection details are  expanded in (c). For $\beta<-1$ there is one intersection with the $h$ axis and the situation is similar to that described in (2a); there is however one important difference in that the curve has a minimum for finite $h$. As $\beta$ is increased above $-1$ a new steady state arise from infinity, leading to a pair of steady states. This scenario corresponds to spinodal conditions reached for $\beta=-1$. The pair of steady states exists in a finite range for $-1<\beta<\beta_{c}$ the value of $\beta_{c}$ being non universal and depending on $\alpha$. In this domain, depending on initial conditions one can have a stable finite thickness or proliferation for $v_{\Lambda}>0$, a stable thickness or collapse if $v_{\Lambda}<0$. So $\beta_c$ marks a discontinuous transition to tissue 
proliferation (resp. tissue collapse) for $v_{\Lambda}>0$ (resp. $v_{\Lambda}<0$). 
 (d) the three curves corresponding to  $\beta=-2,-1$ and $0$ are generic for  
$\alpha>1$. For all $\beta< -1$ there is no fixed points and the epithelium either proliferates or collapses depending on the $v_{\Lambda}$ sign; for all $\beta> -1$ there is one fixed point, unstable if  $v_{\Lambda}>0$, stable if  $v_{\Lambda}<0$. The steady state thickness diverges as $\beta \to -1$. This defines the transition at $\beta= -1$ as a continuous transition.}
\label{conttrans}
\end{figure}

\begin{figure}[htb]
\includegraphics[height=8cm]{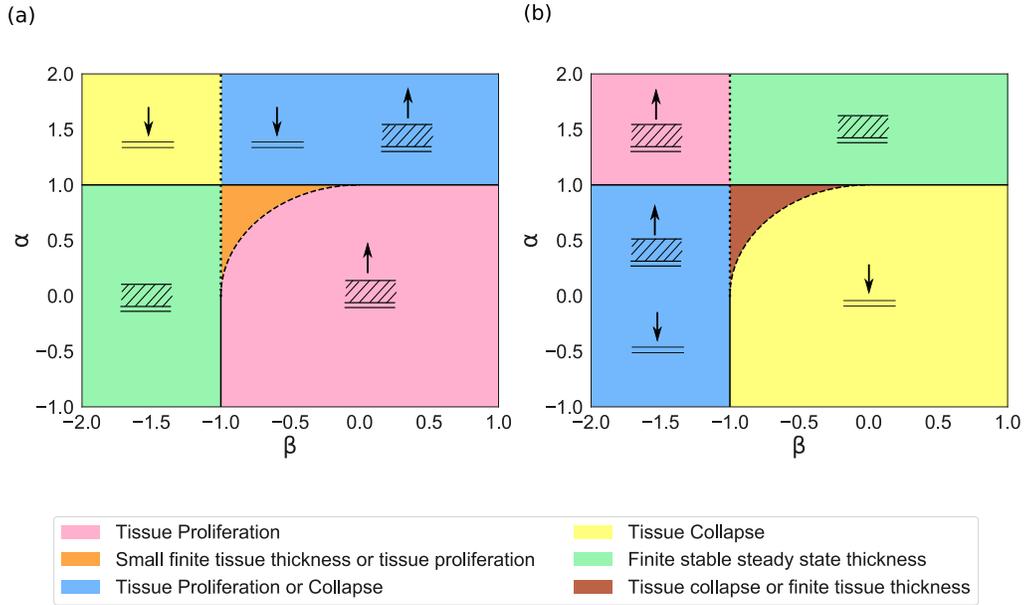}
\caption{
State diagram of a thick epithelium 
in the $\alpha$-$\beta$ space for (a) $v_\Lambda>0$, and (b) $v_\Lambda<0$. In the green region the epithelium layer evolves spontaneously to a finite stable thickness, in the pink region it proliferates, in the yellow region it collapses. In the blue region either it proliferates or collapses depending on initial conditions. In the orange region it can either evolve toward a finite tissue 
thickness or toward proliferation depending on the initial thickness whereas in the brown region it either evolves toward 
tissue collapse or a finite tissue thickness again depending on initial tissue thickness. A continuous black 
line denotes continuous transitions, dashed lines discontinuous transitions, and dotted line 
spinodals.}
\label{phaseab}
\end{figure}

 We show in Fig. 2 
 examples of the three generic scenarios 
characterizing the dependence of $dh/dt$ on tissue thickness $h$.
 Zeros of f(h) 
  correspond to steady states which can be stable or unstable depending on the
 sign of the slope $df/dh$. Negative slopes correspond to stable and positive slopes to unstable situations. 
 We find that either two, one or zero steady states may exist, depending on the values of $\alpha$ and 
 $\beta$. 
 The different possible scenarios are summed up in two 
 state diagrams shown in Fig. 3 (a) and (b), for 
 $v_\Lambda>0$, and $v_\Lambda<0$ respectively.  
 The lines delineating the 
  states characterise 
   continuous and discontinuous transitions. Spinodal lines are also shown. 
Discontinuous transitions, are reached when the conditions $f(h)=0$ and $df/dh=0$ are satisfied simultaneously. 
The corresponding line is a portion of the circle $\alpha^2+\beta^2=1$ as shown on Fig.2. It will be shown as a dashed line on all figures. All other boundaries correspond  either to continuous transitions shown as 
solid lines, or to `spinodal' transitions shown as dotted lines. The `spinodal' 
lines signal the appearance/disappearance of an unstable (resp: stable) steady state, when crossing the line from a phase in which only one stable (resp: unstable) steady state exists. In the corresponding domain, initial conditions determine the eventual fate of the system.

As the coefficients $\alpha$ and $\beta$ depend on the parameters defining the 
  tissue properties and experimental conditions, we 
can transform the generic state diagrams in diagrams corresponding to conditions in which the external flow $v^f_{\rm ext}$ or the current $I_{\rm ext}$ are imposed. We can also investigate the role of parameters such as
the effective homeostatic pressure $\bar P^H$.
 We discuss the steady states and the different growth regimes 
  as a function of
$v^f_{\rm ext}$ and $I_{\rm ext}$ in the next subsection.

 \subsubsection{Steady states}
 
When a steady state 
with $dh/dt=0$ in Eq. (\ref{dhdt2}) exists,  the steady state tissue thickness is given by 
\bea
h=\lambda_0\ln\left[{-B \pm \sqrt{B^2-4AC} \over 2A}\right], \label{ssthick}
\eea
This thickness is of order the screening length $\lambda_0$ up to a 
 logarithmic factor. Here
\bea
A &=& \alpha\left[1- {\lambda_0\nu_{\mathrm{eff}}\bar\kappa \over \eta_{\mathrm{eff}}(1-\phi)\Lambda}
- {\lambda_0 \over \eta_{\mathrm{eff}}\Lambda^f}\right]+\gamma_1+\gamma_2 +\beta\left[1 
+ {\lambda_0\nu_{\mathrm{eff}}\kappa_{\mathrm{eff}} \over \lambda(1-\phi)
\eta_{\mathrm{eff}}}\right], \\
B &=& 2(1-\alpha-\beta-\gamma_1), \\
C &=& \alpha\left[1+ {\lambda_0\nu_{\mathrm{eff}}\bar\kappa \over \eta_{\mathrm{eff}}(1-\phi)\Lambda}
+ {\lambda_0 \over \eta_{\mathrm{eff}}\Lambda^f}\right]+\gamma_1-\gamma_2 +\beta\left[1 
- {\lambda_0\nu_{\mathrm{eff}}\kappa_{\mathrm{eff}} \over \lambda(1-\phi)
\eta_{\mathrm{eff}}}\right]. \label{c}
\eea
Where we have introduced 
the dimensionless quantities $\tilde v^f=v^f_\mathrm{ext}/v_2$, 
$\tilde I_{\rm ext}=\lambda I_{\rm ext}(1-\phi)/
(\Lambda \kappa_{\mathrm{eff}}v_2)$, $\gamma_1=\hat v_\Lambda/v_2$,  $\gamma_2=\lambda_0(\bar P^H 
- \nu_{\mathrm{eff}} I_{\rm ext}
/\Lambda + \nu_{\mathrm{eff}}\bar\kappa v^f_\mathrm{ext}/((1-\phi)\Lambda)+ v^f_\mathrm{ext}/
\Lambda^f)/(\eta_{\mathrm{eff}}v_2)$, 
where $\hat v_\Lambda=v_\Lambda - v^f_\mathrm{ext} - 
\lambda I_{\rm ext}(1-\phi)/(\Lambda \kappa_{\mathrm{eff}})$.

Note that steady states exist only if $A>0$, and $-B\pm \sqrt{B^2-4AC}>2A$ or  
if $A<0$ and $-B\pm \sqrt{B^2-4AC}<2A$. 


\subsubsection{Tissue dynamics in the absence of electric currents}
\label{noelec}

We first discuss examples of thickness dynamics in the presence of fluid flow
and absence of electric currents, $I_{\rm ext}=0$. 
Figure 4 shows the 
growth rate $dh/dt$ as a function of tissue thickness for 
$v_2>0$, i.e. increased cell division rate at the tissue-substrate interface. There are three possible behaviours: (i) 
In Fig. (4a) the existence of a stable steady state shows that under the chosen conditions the epithelium layer will spontaneously evolve toward a finite thickness irrespective of the initial conditions; with the chosen numbers this thickness is a fraction of the screening length $\lambda_{0}$, (ii) Fig. (4b), the existence of a pair of stable-unstable fixed points shows that for initial thicknesses smaller than the unstable one the layer goes spontaneously to the stable value whereas for larger initial values the layer proliferates;  
 Fig. (4c), $dh/dt$ being always positive the epithelium layer always proliferates. These different scenarios, depend on the value of the external fluid flow $v^f_{\rm ext}$ and 
tissue homeostatic pressure $P^c_h$. For a negative homeostatic pressure, when cells die in the bulk, a 
positive external fluid flow permeating through the tissue gives rise to cell division and hence 
can counter the effect of cell death. 
When the two effects exactly balance, the tissue exhibits a steady state 
with a finite thickness (Fig. (4a)). 
For a high external fluid flow, cell division surpasses cell apoptosis, and the tissue grows indefinitely (Fig. (4c)). For intermediate values of $v^f_{\rm ext}$, 
the tissue can exhibit 
two different behaviours  depending on 
the initial tissue thickness: it can either reach a stable steady state or proliferate out of bounds (Fig 4b).

\begin{figure}[htb]
\includegraphics[width=5.2cm,height=4.5cm]{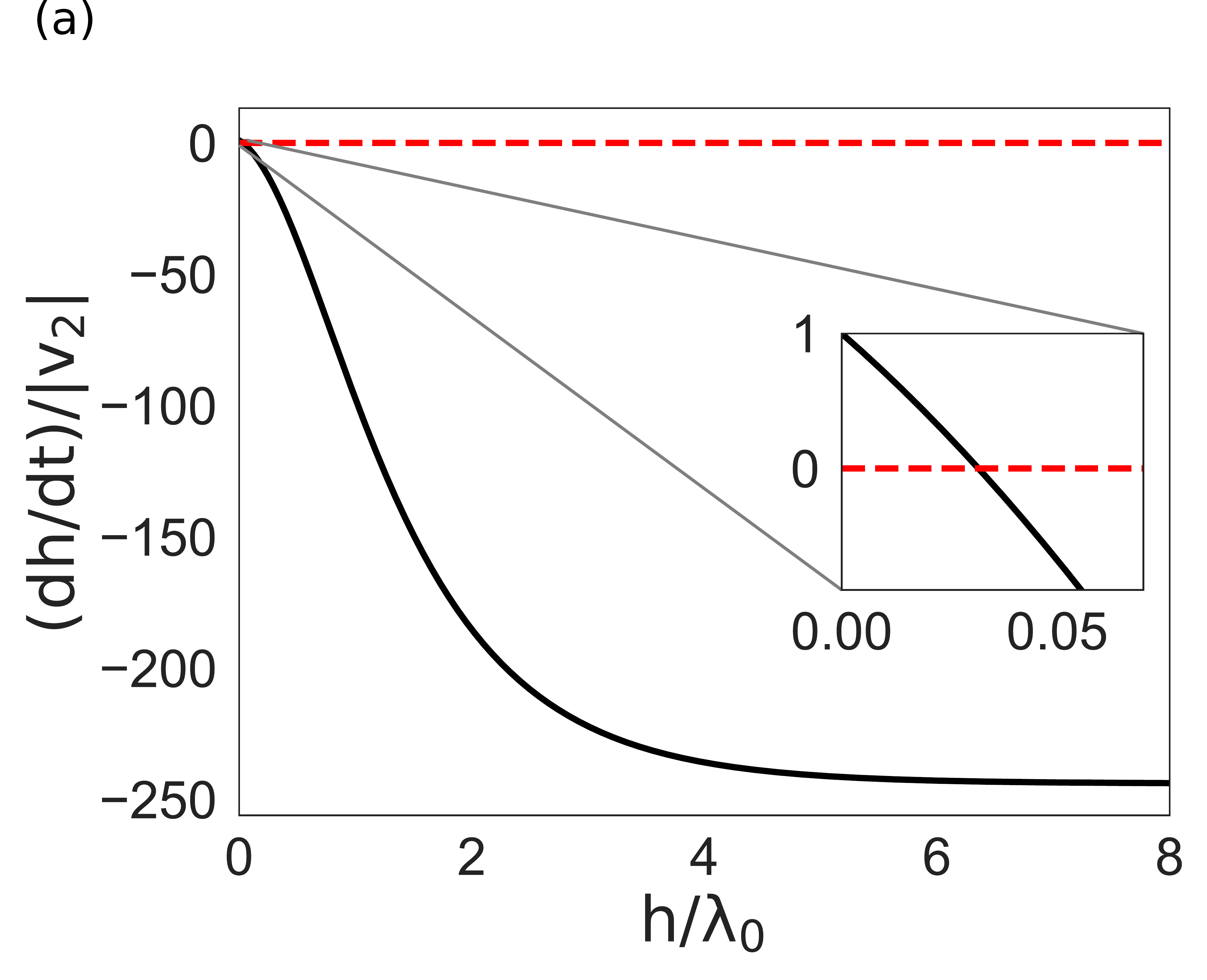}
\includegraphics[width=5.2cm,height=4.5cm]{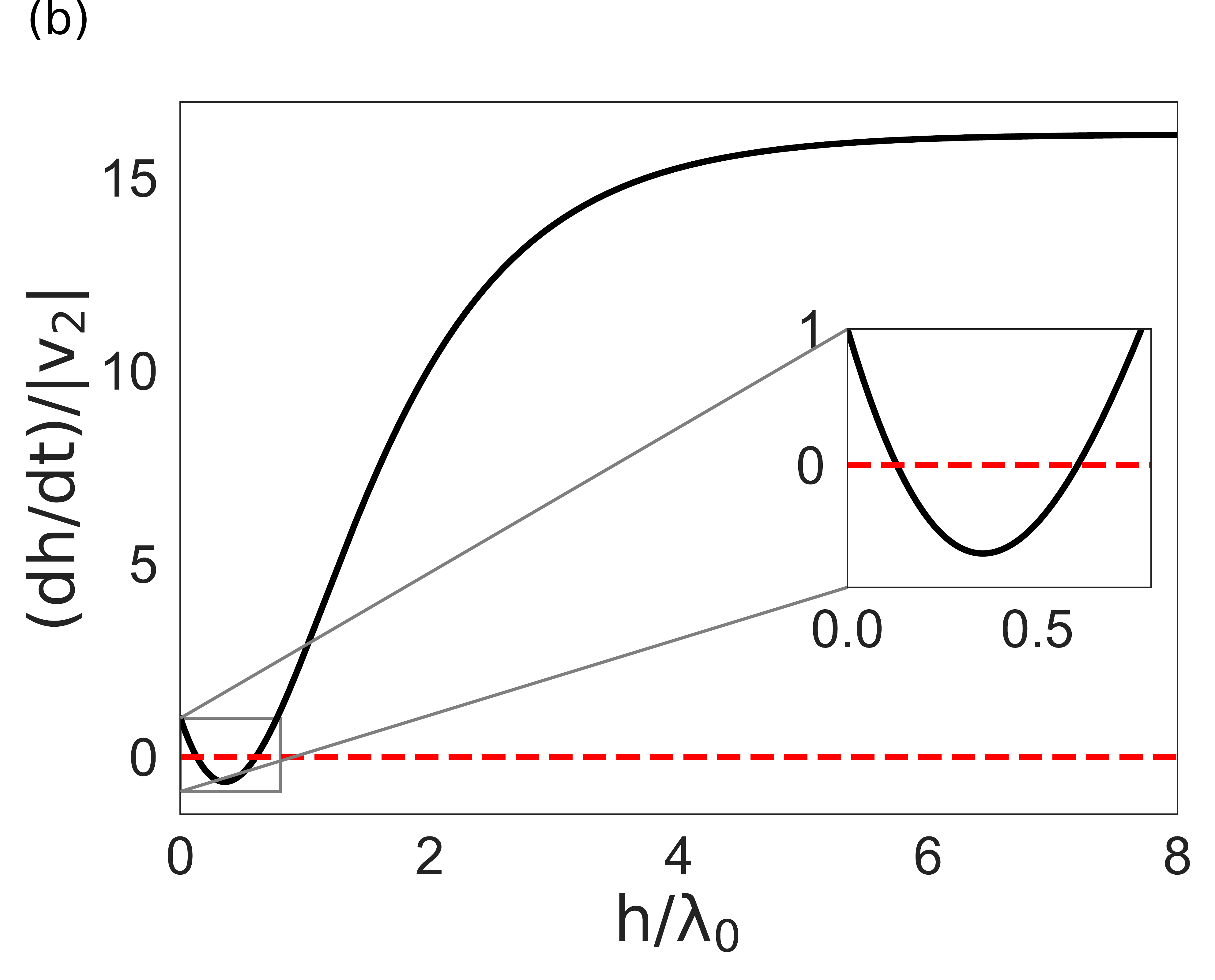}
\includegraphics[width=5.2cm,height=4.5cm]{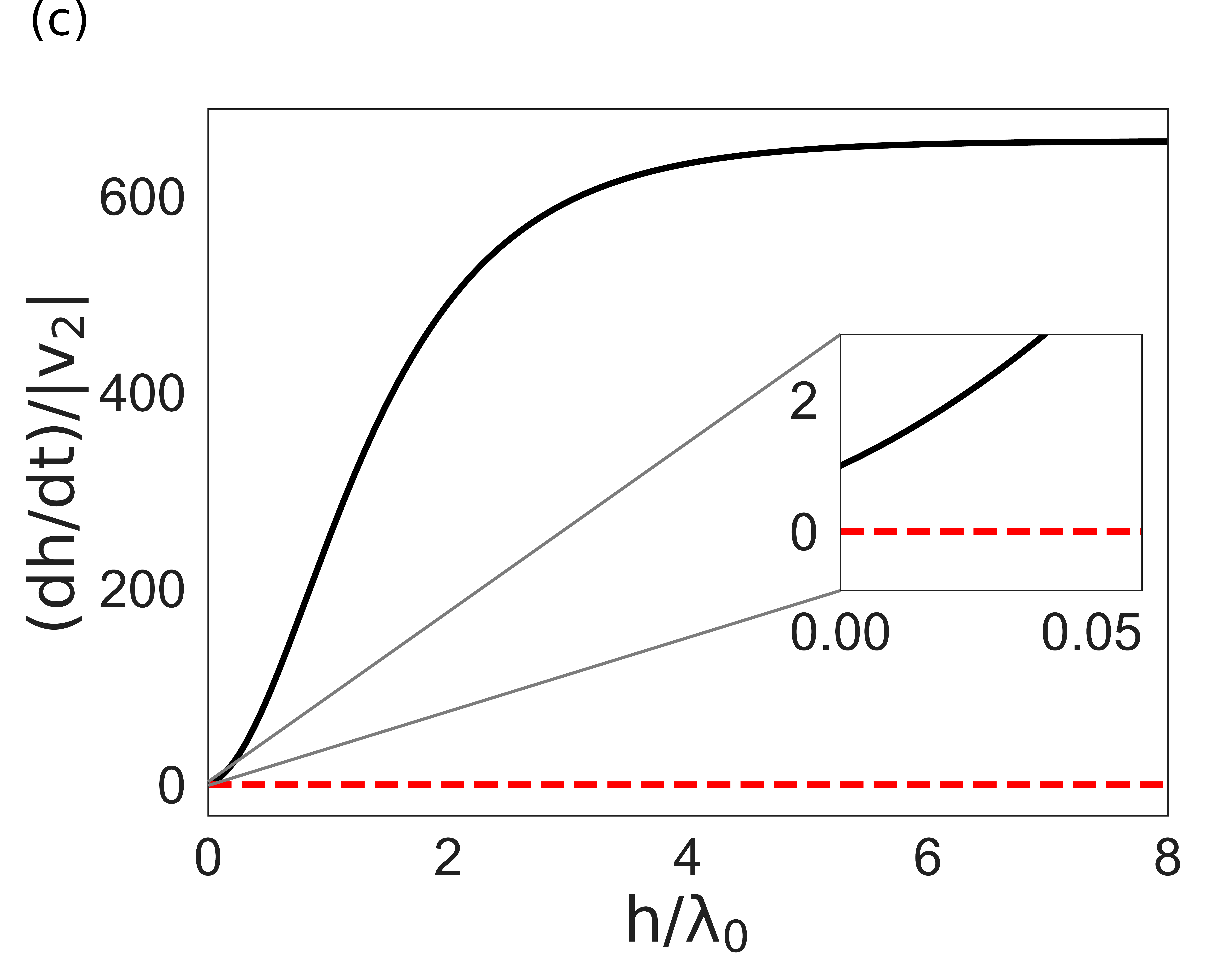}
\caption{
Plots of $dh/dt$ vs $h$ for $v_2>0$ ($v_2=3 \times 10^{-10} {\rm m/s}$), corresponding to an increased
cell division rate,  for negative homeostatic pressure $P^c_h=-15$KPa, showing (a) stable 
steady state resulting in a finite layer thickness for 
$v^f_{\rm ext}=3 \times 10^{-8} {\rm m/s}$, (b) 
a pair of stable-unstable steady states resulting in either finite thickness or proliferation depending on initial thickness
 for $v^f_{\rm ext}=1.08 \times 10^{-7} {\rm m/s}$, and (c) 
absence of any steady state resulting in proliferation for $v^f_{\rm ext}=3 \times 10^{-7} {\rm m/s}$. 
All other parameter values are taken from Table \ref{tab1}.}
\label{thicksimp}
\end{figure}

\begin{figure}[htb]
\includegraphics[width=7cm,height=6cm]{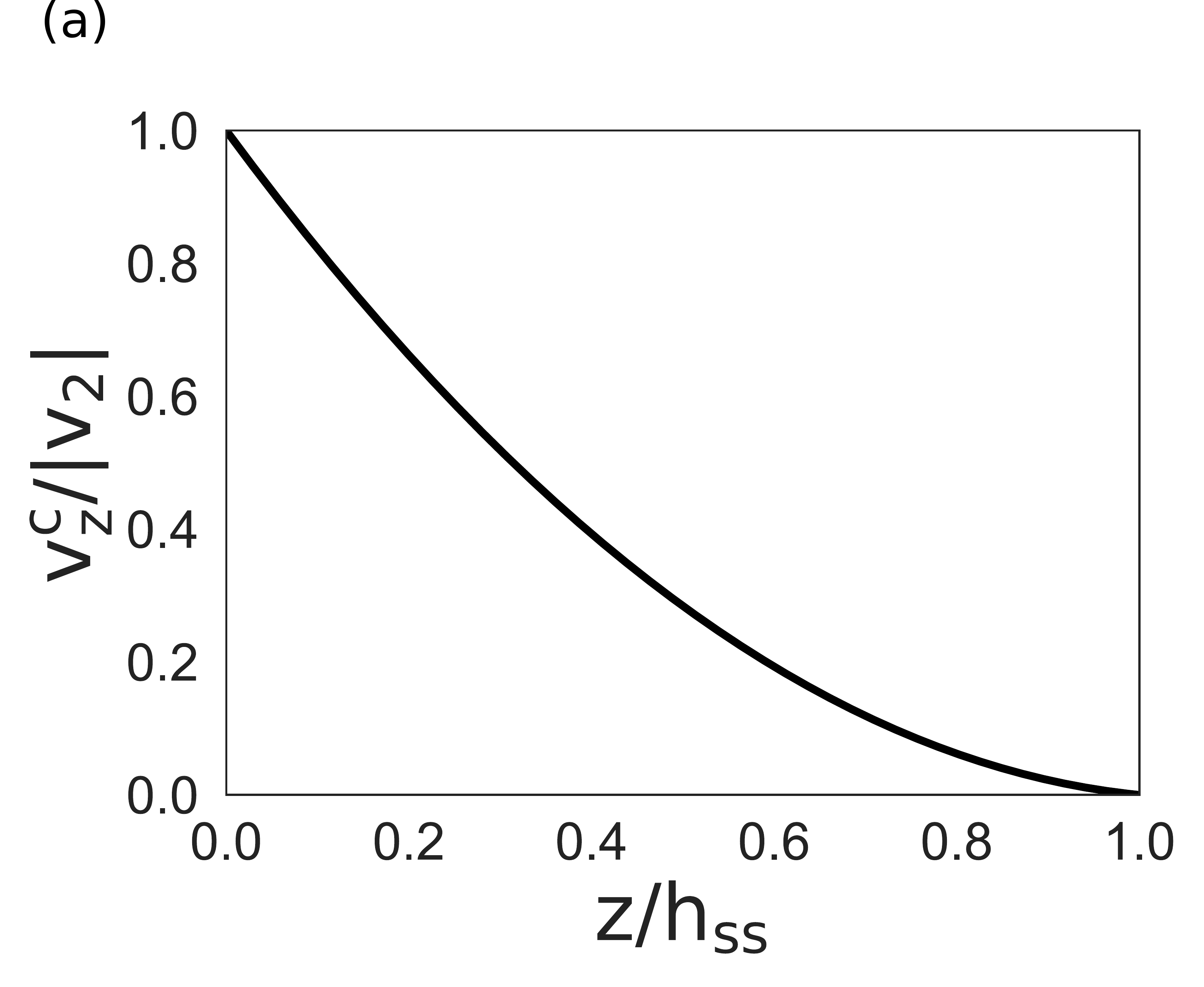}
\includegraphics[width=7cm,height=6cm]{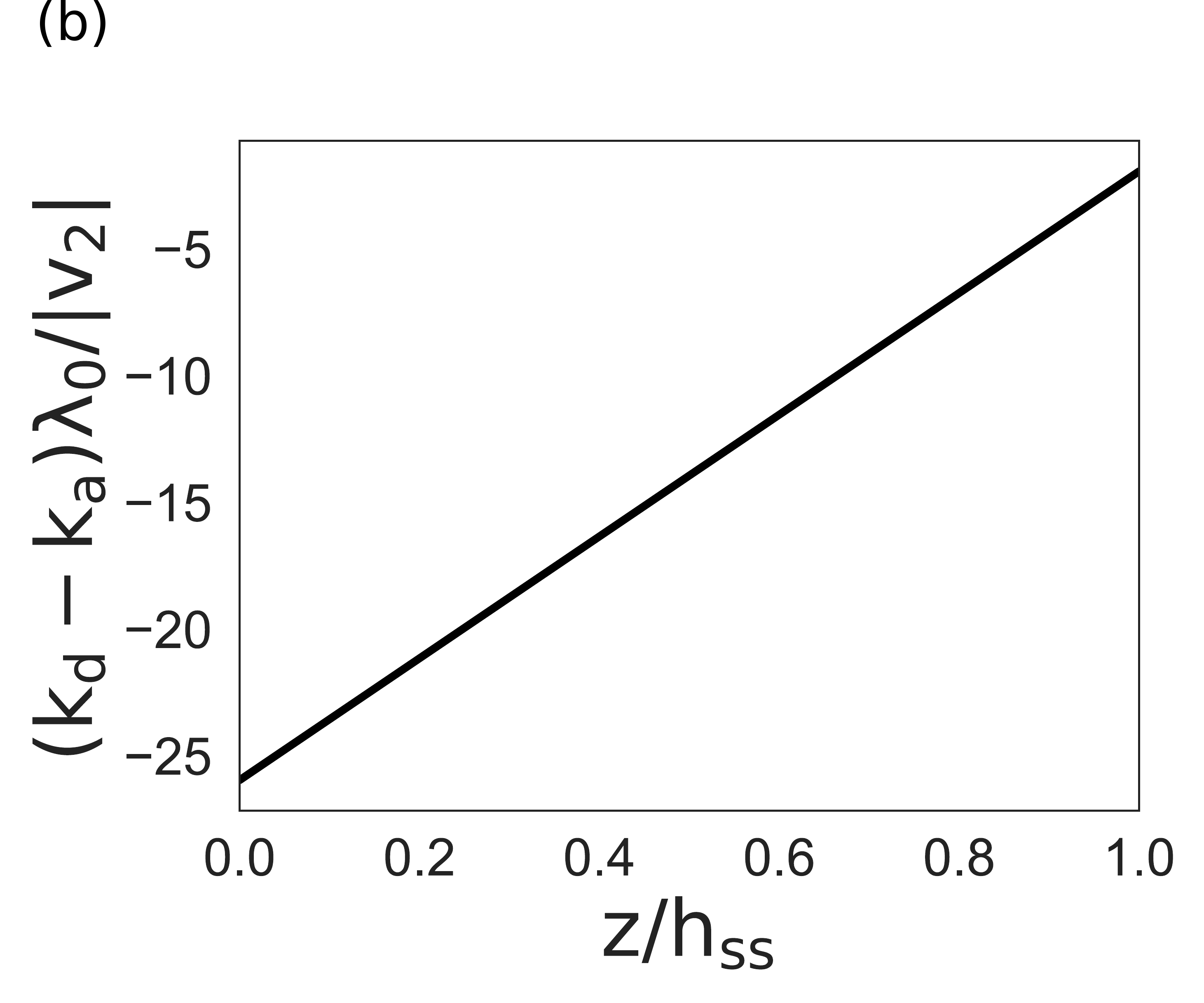}
\includegraphics[width=7cm,height=6cm]{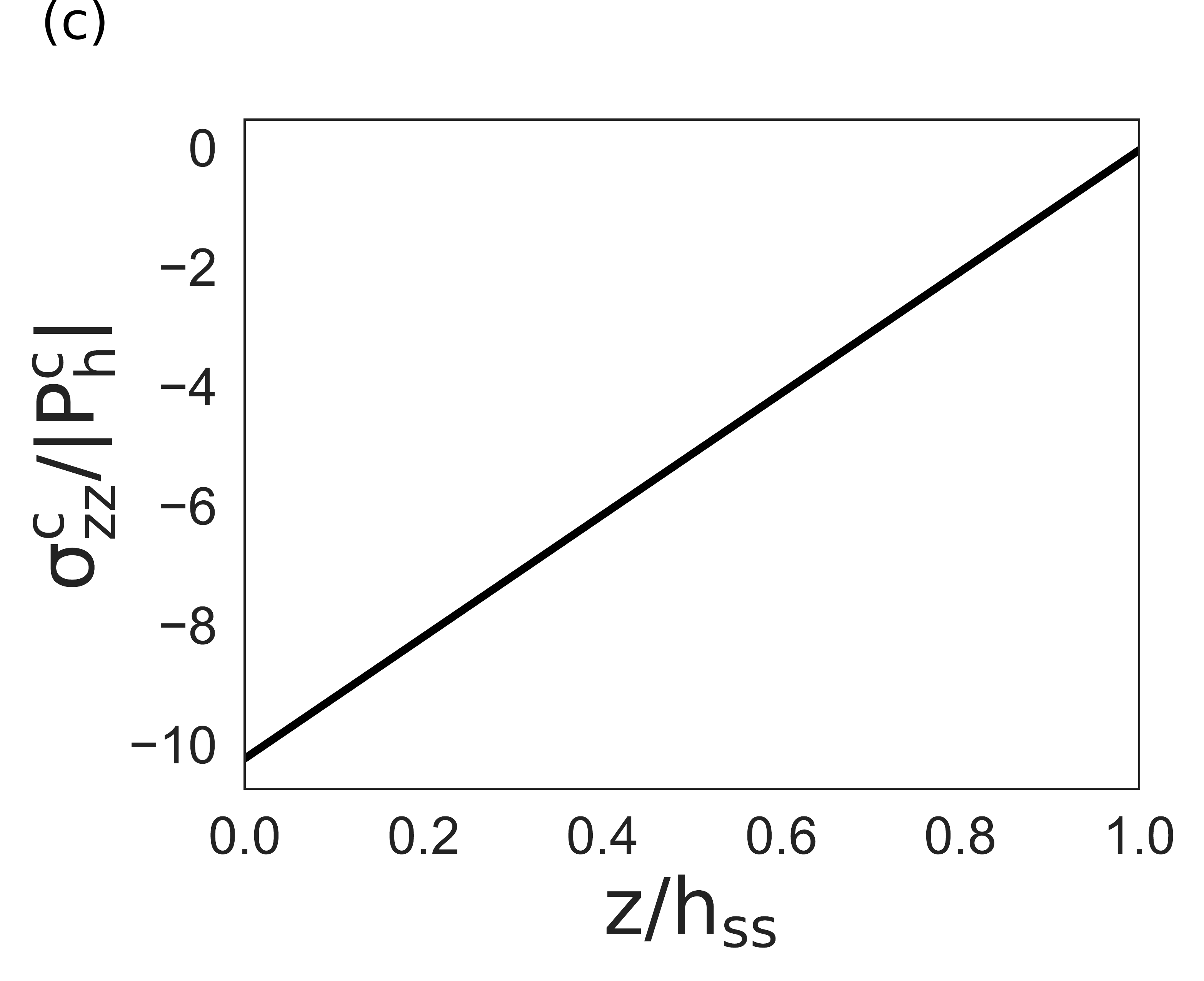}
\includegraphics[width=7cm,height=6cm]{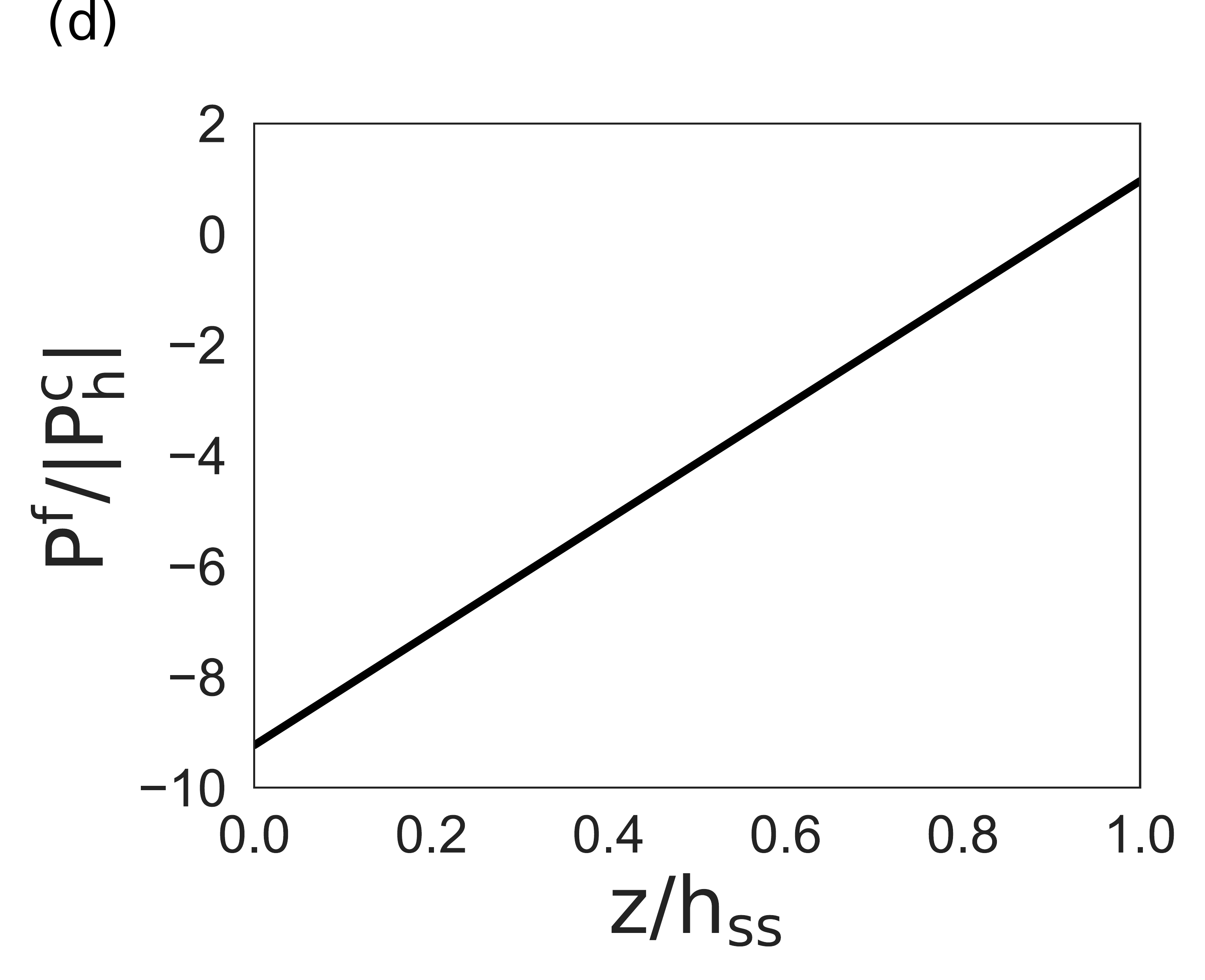}
\caption{Profiles of (a) cell velocity, (b) net cell division rate, (c) total 
cell stress, and (d) fluid pressure in steady states of a thick epithelium  for $v^f_{\rm ext} = -3 \cdot 10^{-10} \mathrm{m/s}$,  $I_{\rm ext}=0$, $P^c_h=-1$ KPa and $v_2=3 \cdot 10^{-10} \mathrm{m/s}$.
  Other parameter values are taken from Table \ref{tab1}.}
\label{ssprosimp}
\end{figure}

At steady state, the tissue slab is not homogeneous. The net cell 
turnover, the cell velocity, the total cell stress, and the fluid 
pressure exhibit non-trivial profiles along the $z$-axis. An example of profiles is 
shown in Fig. (\ref{ssprosimp}) for $v^f_{\rm ext} = -3 \cdot 10^{-10} \mathrm{m/s}$ and $I_{\rm ext}=0$ 
and homeostatic pressure of $P^c_h=-1$ 
KPa. In this case cells undergo apoptosis in the bulk, and only the cell division in the surface layer  
prevents the tissue from collapsing. 
The fluid flowing in the 
negative $z$ direction, translates into a pressure on the tissue which increases the rate of 
apoptosis. This effect is cumulative and the pressure is larger at the bottom of the layer than at the top, as can be seen on the stress profile. At the bottom layer, cells divide; as they move up, they balance cell death. As required by boundary conditions the cell velocity vanishes at the free surface. Since the total apoptosis rate is larger in the presence of flow than in the absence of flow, the steady state thickness is smaller in the presence of flow than in its absence.

\begin{figure}[htb]
\includegraphics[width=5.2cm,height=4.5cm]{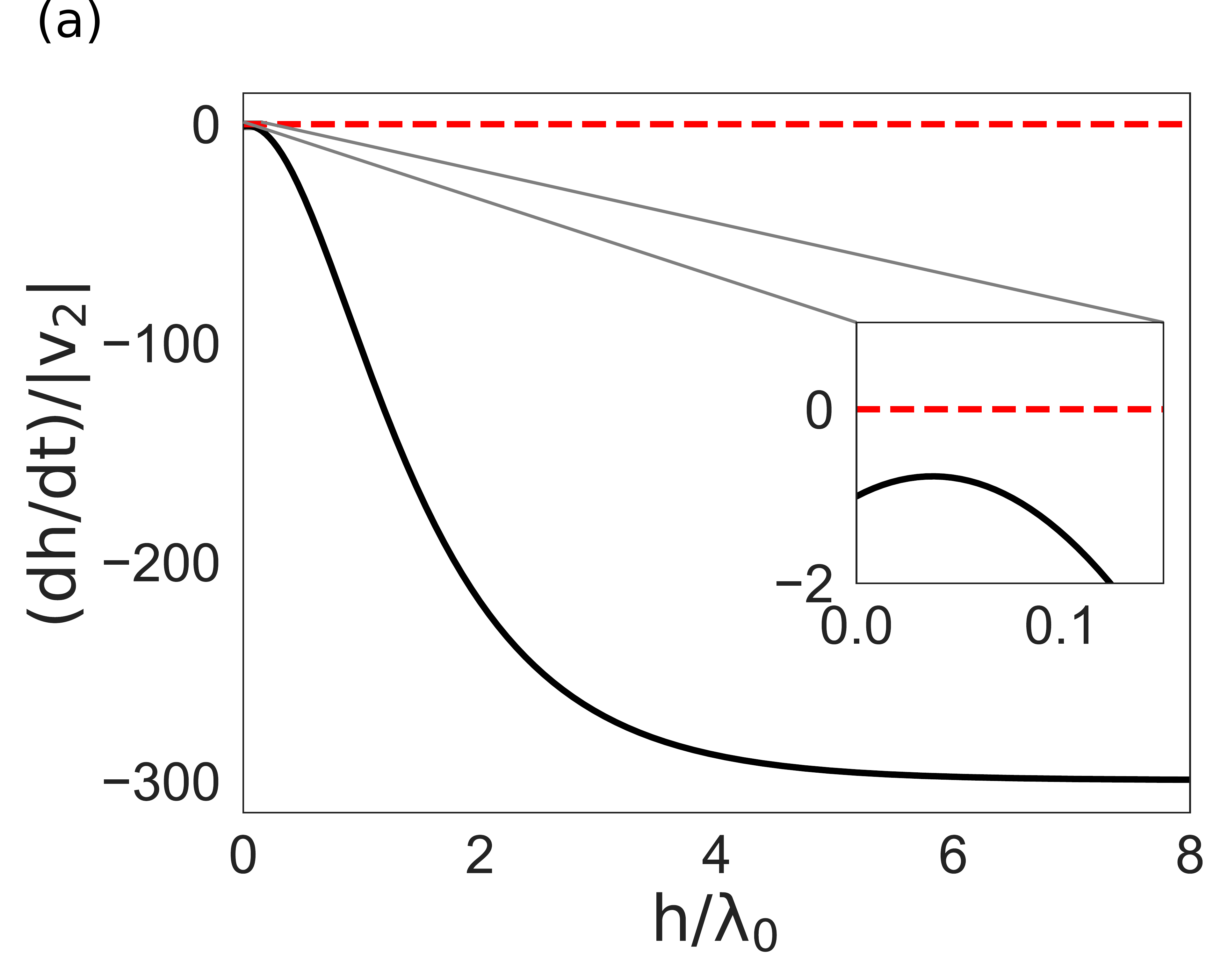}
\includegraphics[width=5.2cm,height=4.5cm]{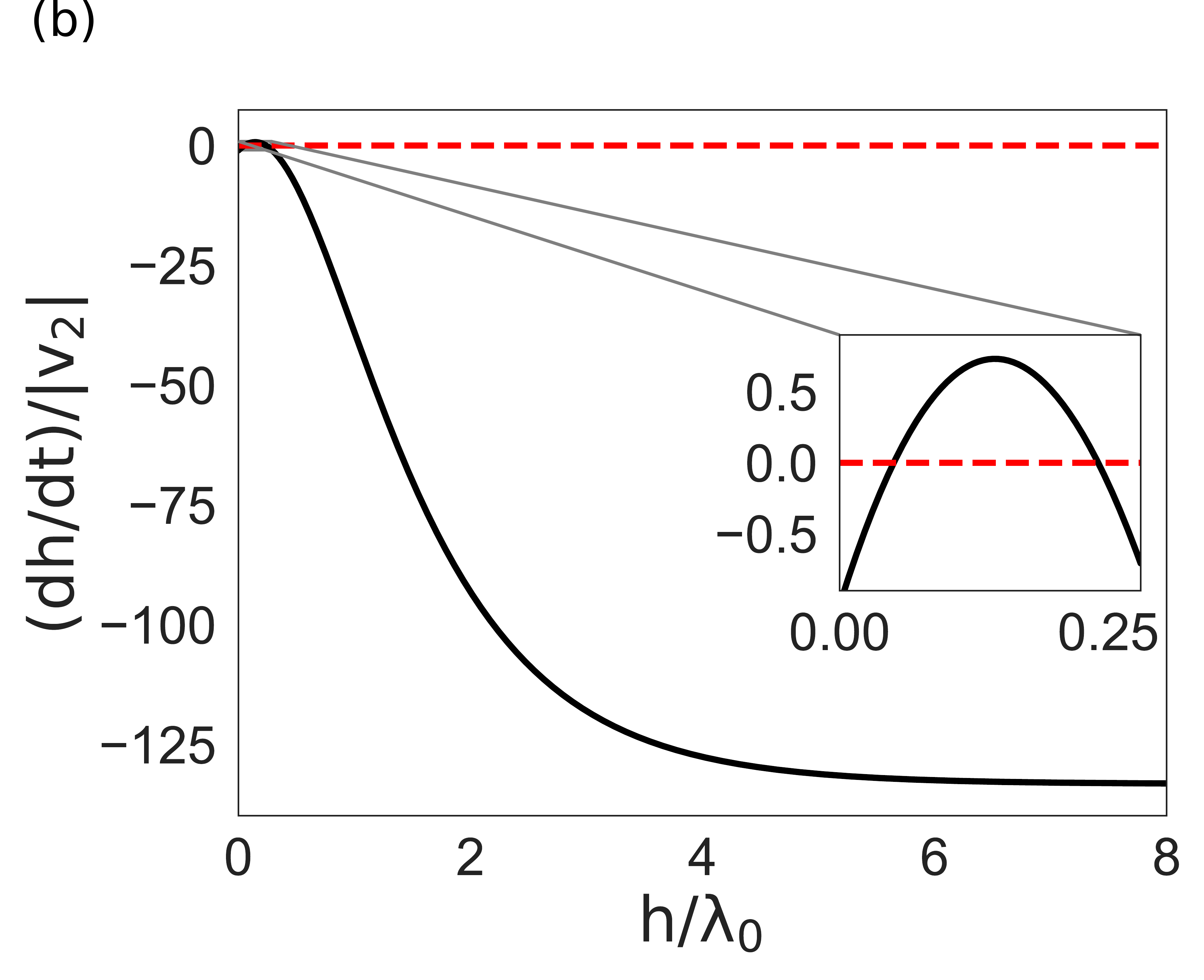}
\includegraphics[width=5.2cm,height=4.5cm]{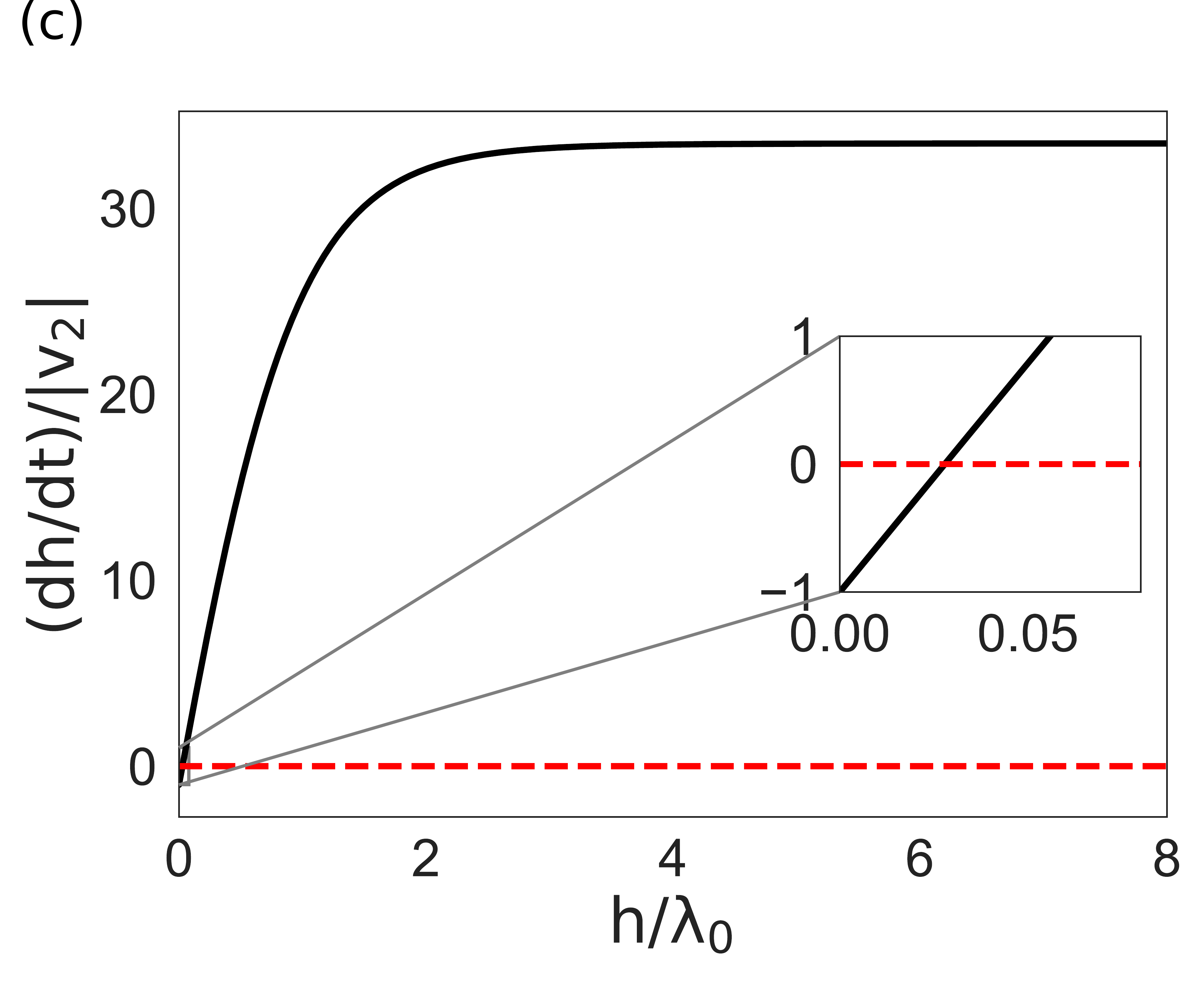}
\caption{
Plots of $dh/dt$ vs $h$ for $v_2<0$ ($v_2=-3 \times 10^{-10} {\rm m/s}$) at $P^c_h=5$KPa showing (a) absence of steady state resulting in layer collapse 
 for 
$v^f_{\rm ext}=1 \times 10^{-10} {\rm m/s}$, (b) presence of a pair of unstable-stable steady stattes resulting in either a layer collapse or a finite stable thickness depending on the initial thickness value, 
for $v^f_{\rm ext}=5 \times 10^{-8} {\rm m/s}$, and (c) presence of an unstable steady state resulting in either layer collapse or proliferation depending on initial thickness value for 
$v^f_{\rm ext}=1 \times 10^{-7} {\rm m/s}$. 
All other parameter values are taken from Table \ref{tab1}.}
\label{thicksimp2}
\end{figure}

The dynamics of the tissue for $v_2<0$, i.e., cells dying at the tissue-substrate interface, is shown in Fig. 
(\ref{thicksimp2}). There are three possible scenarios, 
 which 
  are again determined by the values of $v^f_{\rm ext}$, and $P^c_h$. For $P^c_h>0$, and a low 
$v^f_{\rm ext}$, cell death still dominates, and the tissue collapses fully as shown in Fig. (\ref{thicksimp2}a). For $P^c_h>0$, 
and intermediate values of $v^f_{\rm ext}$, 
the tissue can either reach a stable thickness or collapse  
depending on the initial  thickness. This possibility is shown in Fig. (\ref{thicksimp2}b). 
Eventually, for large enough $v^f_{\rm ext}$, the steady state thickness regime disappears in favour of proliferation: depending on the initial thickness, the tissue either collapses or proliferates. This unstable regime is shown in Fig. (\ref{thicksimp2}c).

State diagrams 
 in the $\tilde P^H$ - $\tilde v^f$ phase space 
 summarize these results in Fig. (
7a) for $v_2>0$ and 
Fig. (
7b) for $v_2<0$. Here $\tilde P^H=\lambda_0 P^c_h/(\eta_{\rm eff} v_2)$. Changing 
$\tilde P^H$ 
can be achieved by tuning $P^c_h$ and, changing $\tilde v^f$ 
by tuning $v^f_{\rm ext}$. 
 As already explained for $v_2>0$, the tissue 
can exhibit three different behaviours, depending on the 
values of $\tilde P^H$ and $\tilde v^f$: (i) the green region corresponds to the existence of a stable steady state with a finite  thickness irrespective of initial thickness, (ii) 
the orange region corresponds to  tissue evolving toward a finite thickness or proliferating  depending on the initial  thickness, and, (iii) the pink region corresponds to uncontrolled growth. 
The solid line delineating the green and the pink regions indicates a continuous transition: the steady state thickness diverges upon approaching the line and reaches infinity on the line. The dotted line delineating the green and orange regions correspond to a spinodal line. It signals the appearance of an unstable fixed point at finite thickness simultaneously with the appearance of a stable fixed point for infinite thickness when crossing the line from green to orange. The dashed line which delineates the orange and pink regions is a line 
of discontinuous transition, where the pair of stable-unstable fixed points with finite thickness disappears when crossing the line from orange to pink.

A negative $v_2$ implies a sink for cells at the tissue-substrate boundary. 
There are again three possible scenarios: (i) in the blue region, one has one unstable steady-state: for initial thicknesses smaller than that of the unstable fixed point the tissue collapses, while for larger initial thicknesses  the tissue proliferates, (ii)  in the brown region, there are two fixed points, one unstable and one stable: the tissue collapses for initial thicknesses smaller than that of the unstable fixed point, and converges to a finite stable thickness if the initial thickness is larger than that of the unstable fixed point, (iii) in the yellow region, there is no fixed point, $dh/dt$ is always negative the tissue collapses. The continuous line signals a continuous transition from a scenario with one unstable steady states to the total absence of steady state, the value of the unstable thickness going continuously to infinity upon approaching the line from the blue side. The dotted line between the blue and the brown region signals a 
spinodal transition, with the stable fixed point of the brown region going continuously to infinity upon approaching the blue region. The dashed line  signals a discontinuous transition, the stable unstable pair of fixed points disappearing simultaneously for a finite common thickness. 

\begin{figure}[htb]
\includegraphics[width=16cm,height=8cm]{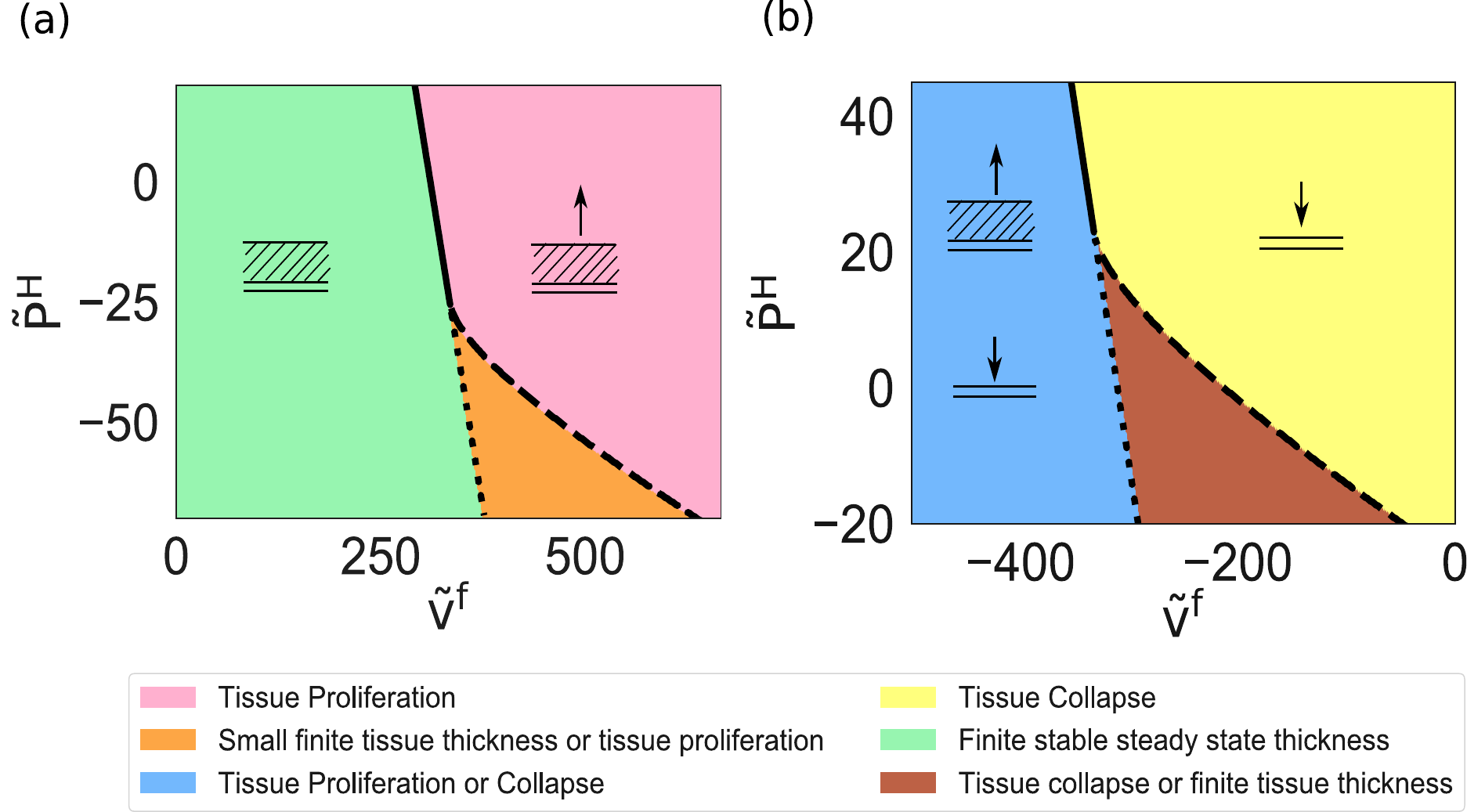}
\caption{State diagram of a thick epithelium in the presence of external fluid flow and absence 
of electric current. 
  (a) for $v_2=3\cdot 10^{-10} {\rm m/s}$ 
   in the green domain the layer evolves spontaneously to a finite stable value, in the pink domain the layer proliferates and in the orange domain it either evolves toward a finite thickness or proliferates depending on initial conditions. b) for  $v_2=-3\cdot 10^{-10} {\rm m/s}$ in the yellow domain the layer collapses, in the brown domain it either collapses or evolve toward a finite stable thickness and in the blue region it either collapses or proliferate depending on initial conditions. Continuous, dashed and dotted lines have the same meaning as on figure (3).
 In 
(a) and (b) we have used 
$\lambda_1=-1 \cdot 10^{-8} {\rm N/m^3}$. The other parameters used are given in Table \ref{tab1}.}
\label{phasesimp}
\end{figure}

\subsubsection{Tissue dynamics in the presence of electric currents}

\begin{figure}[htb]
\includegraphics[width=5.2cm,height=4.5cm]{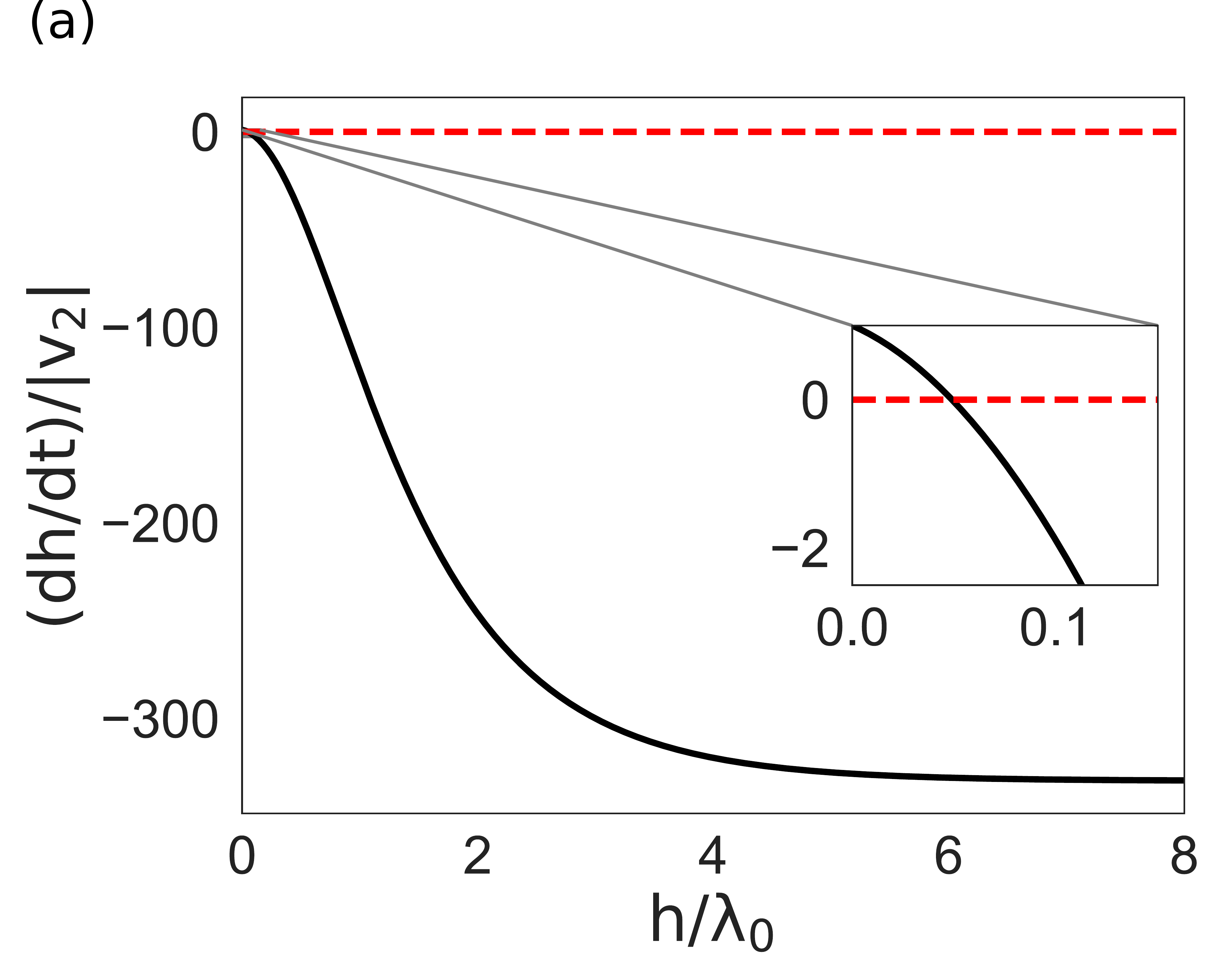}
\includegraphics[width=5.2cm,height=4.5cm]{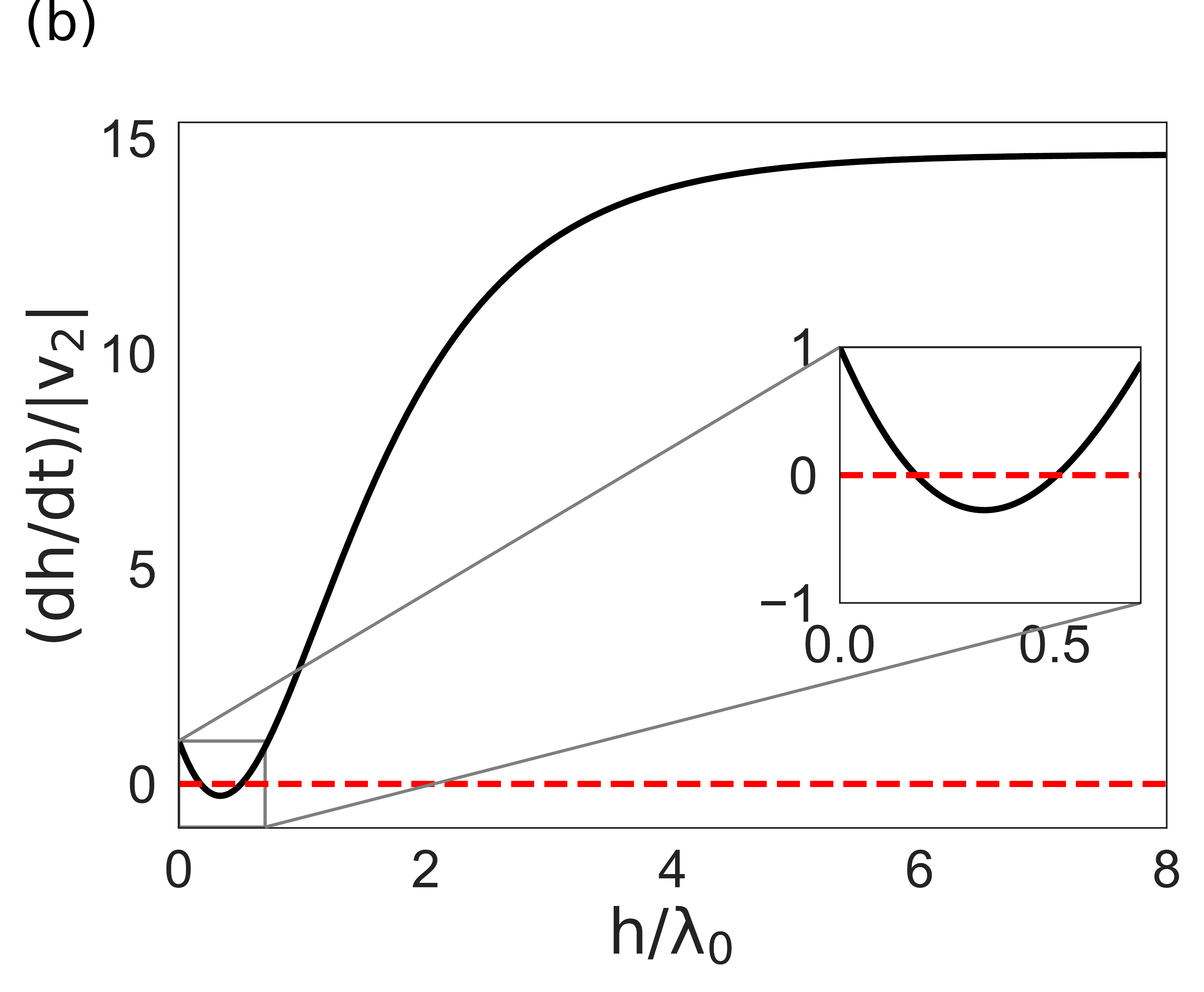}
\includegraphics[width=5.2cm,height=4.5cm]{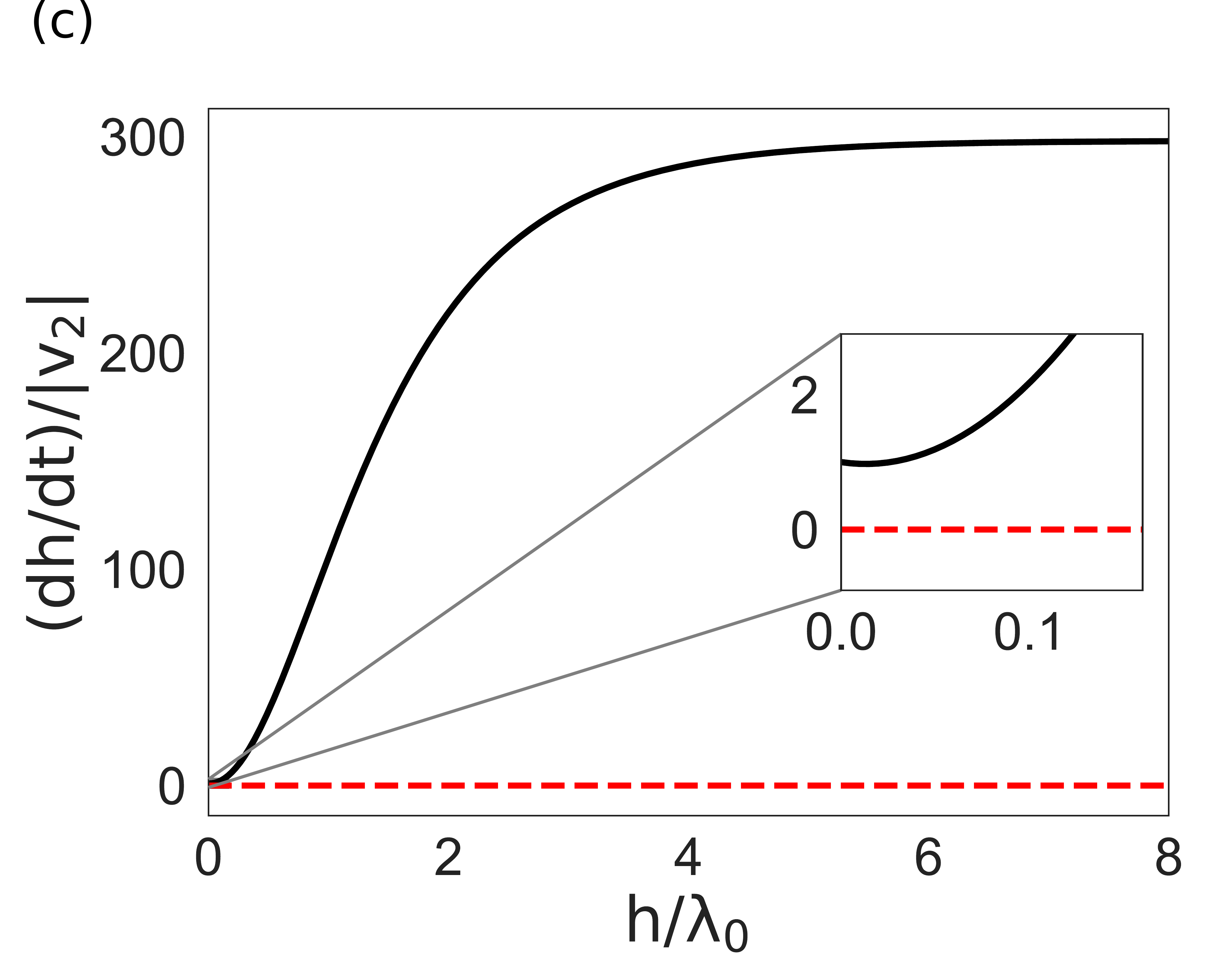}
\caption{
Plots of ${dh/ddt}$ vs ${h}$ for $v_{2}>0$ ($v_2=3 \times 10^{-10} {\rm m/s}$) in the presence of electric current and fluid flow. (a) For a low electric current 
$I_{\mathrm{ext}}=3 {\rm A/m^2}$, we get a 
the layer evolves toward a stable finite thickness 
 (b) For a higher value of the current 
$I_{\mathrm{ext}}=4.1 {\rm A/m^2}$, the layer evolves either toward a stable finite thickness  or proliferates depending on initial thickness. (c) For a high enough value of the electric current 
$I_{\mathrm{ext}}=5 {\rm A/m^2}$, the layer always proliferates. Here we have used 
$P^c_h=-5$KPa and $v^f_{\rm ext}=-3\cdot 10^{-9} {\rm m/s}$. The chosen values of the rest of the parameters are taken from Table \ref{tab1}.}
\label{dhdtphase1}
\end{figure}

We now discuss the state diagram as a function of the imposed electric current $I_{\rm ext}$. 
Using eq. (\ref{dhdt1}) we plot $dh/dt$ vs $h$ in the regime $v_2>0$ in Fig. (\ref{dhdtphase1}), 
for different values of the current $I_{\rm ext}$. 
Like in the previous subsection, we find three 
possible scenarios: (a)  irrespective of its initial thickness, the tissue reaches a stable steady state 
(b) if the initial thickness is larger than the unstable fixed point value the tissue proliferate and if the initial thickness is smaller than this value the tissue thickness goes to a finite stable value 
 (c) $dh/dt$ is always positive and irrespective of initial conditions the tissue proliferates. 
In all these cases $I_{\mathrm{ext}}>0$, 
 the current flows from the substrate towards 
the growing interface of the tissue in the positive $z$ direction; this provides an electric field  promoting tissue growth if $ \nu_{1}>0$. Furthermore for $v_2>0$  the surface layer attached to the substrate acts also as a source of dividing cells. However if the homeostatic pressure, $P^c_h$, and the external fluid velocity 
$v^f_\mathrm{ext}$ are negative, both of which favour cell apoptosis, the tissue slab may reach a stable steady state whenever the opposing effects balance exactly. 
 Such a situation is displayed on Fig. (\ref{dhdtphase1}a), where a balance between the 
positive $I_{\mathrm{ext}}$, and negative $v^f_\mathrm{ext}$ allows for the tissue to reach a steady state. 
In Fig. (\ref{dhdtphase1}b), the value of the 
current $I_{\mathrm{ext}}$ is higher, 
the electric field 
 can no longer be balanced if the initial thickness is large enough and the tissue 
  reaches an uncontrolled growth state. 
However if 
 the initial thickness is small enough the effects of the fields can still be counterbalanced, and the tissue can reach a steady state. 
If the value of $I_{\mathrm{ext}}$ is further increased, as in Fig. (\ref{dhdtphase1}c), the electric 
field is strong enough to push the tissue to the uncontrolled growth phase, where $dh/dt>0$ irrespective of initial conditions.  

\begin{figure}[htb]
\includegraphics[width=7cm,height=6cm]{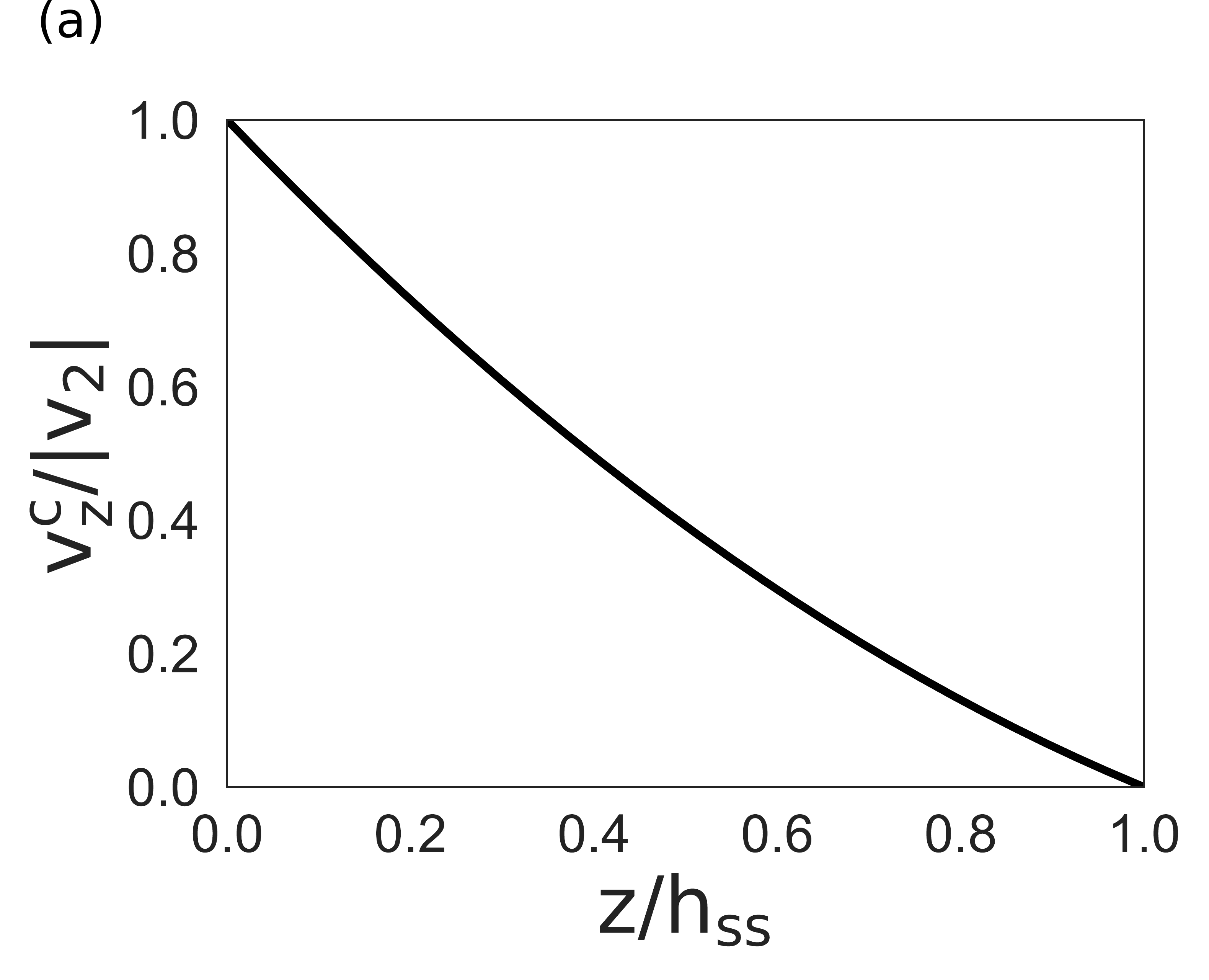}
\includegraphics[width=7cm,height=6cm]{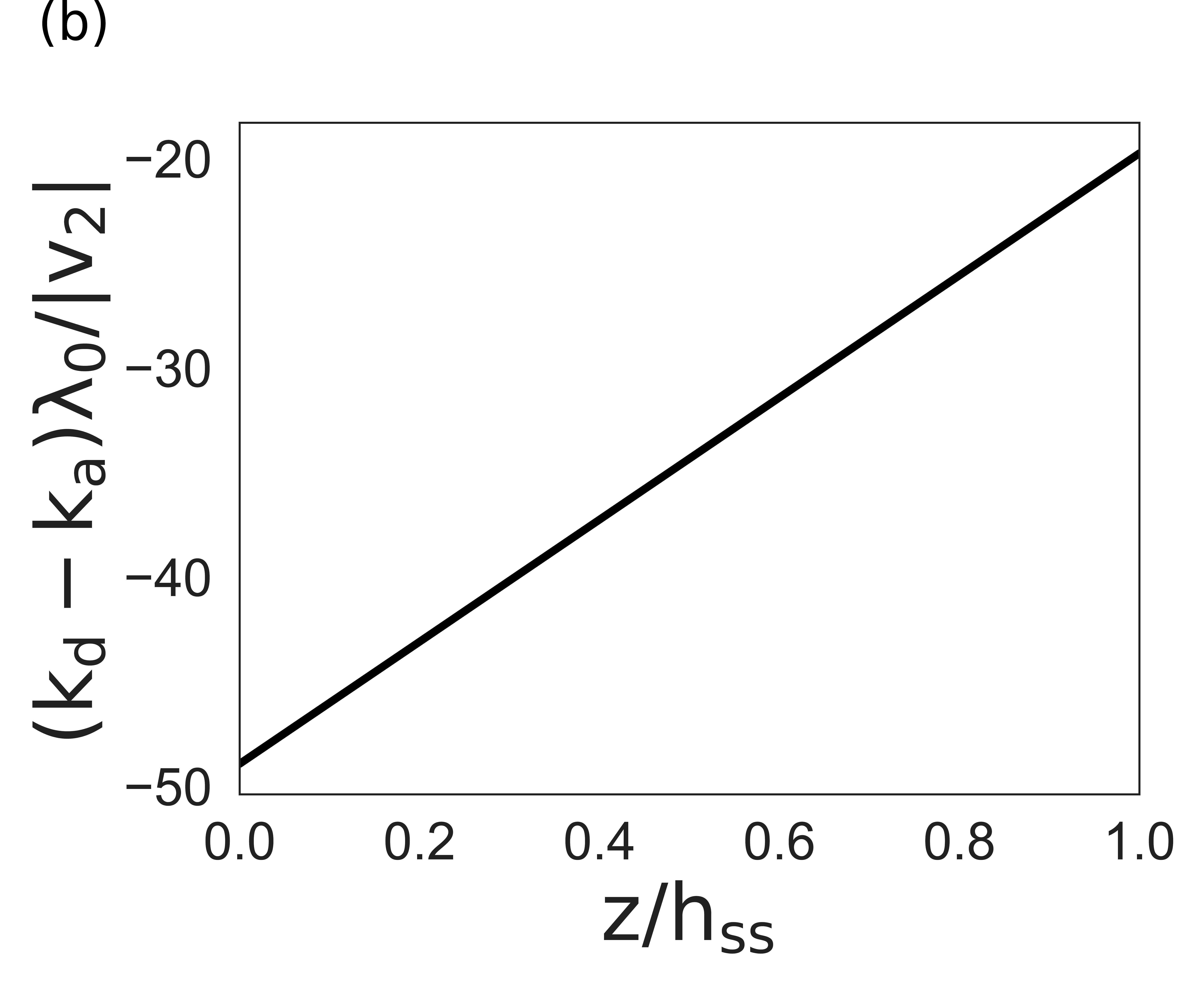}
\includegraphics[width=7cm,height=6cm]{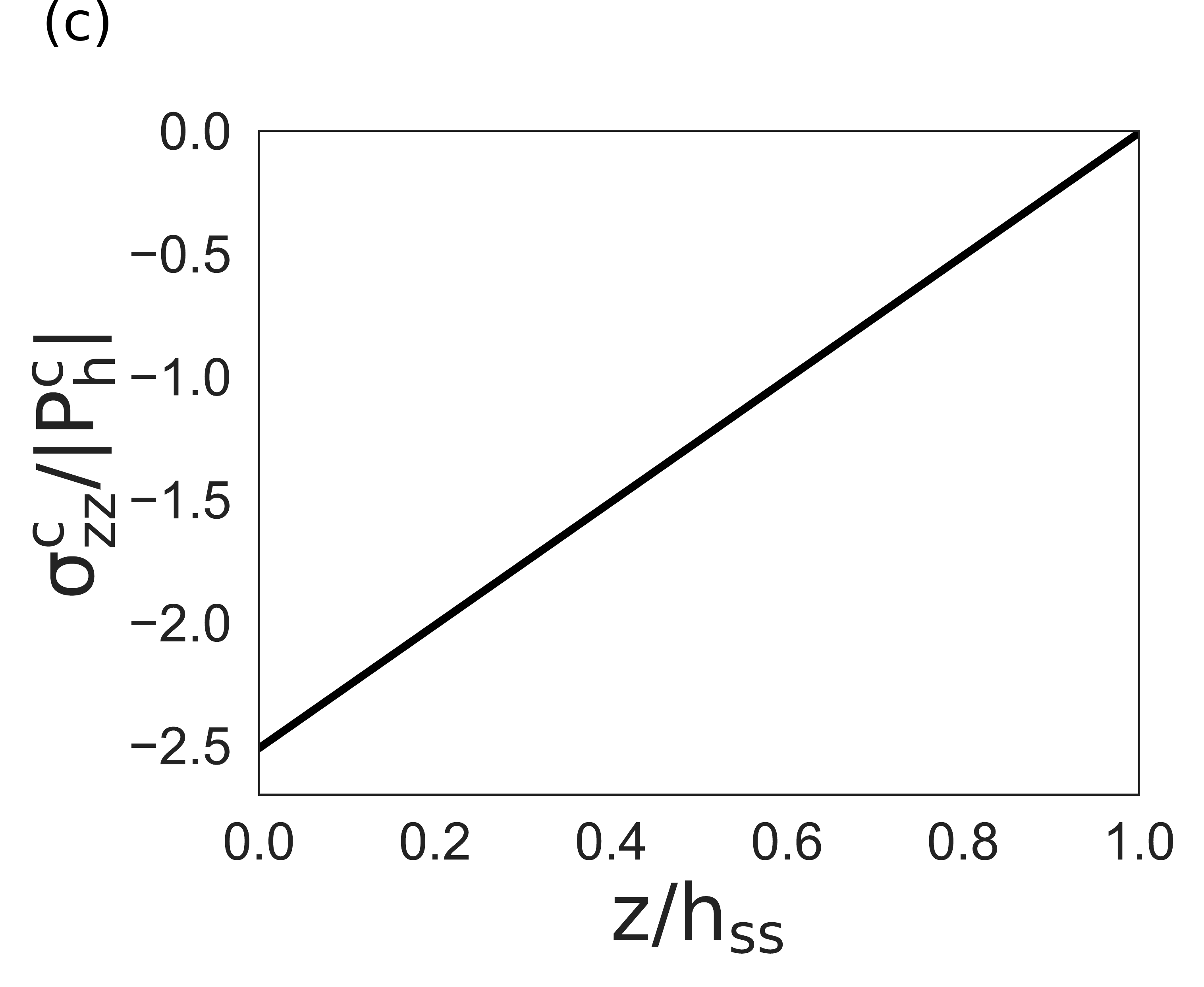}
\includegraphics[width=7cm,height=6cm]{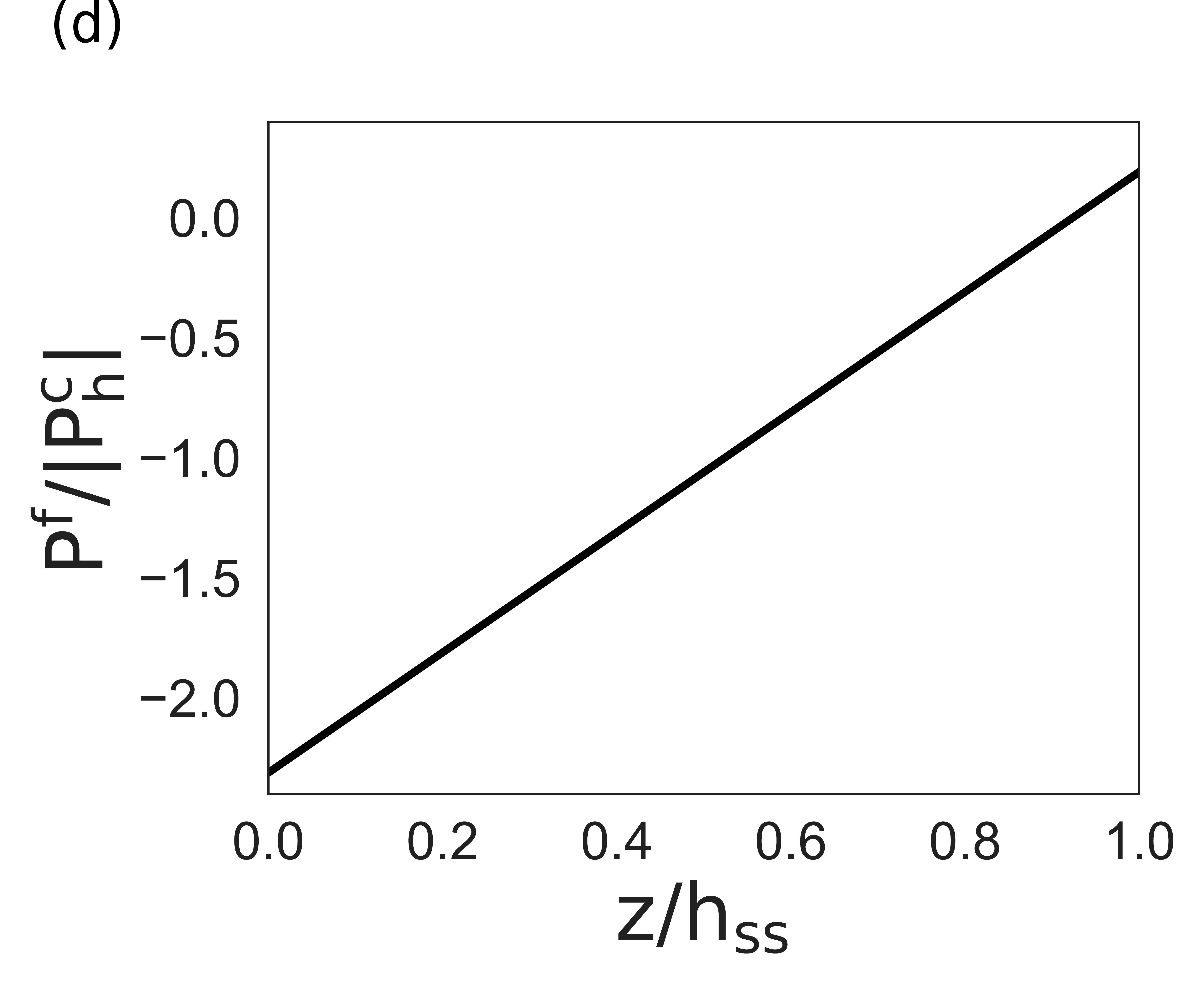}
\caption{Steady states profiles of the 
(a) cell velocity, (b) cell turnover, (c) cell stress and (d) fluid pressure, as a function of the distance to the substrate $z$ for $I_{\mathrm{ext}}=1 {\rm A/m^2}$, $P^c_h=-5 {\rm KPa}$, $v_2=3 \times 10^{-10} {\rm m/s}$, and 
$v^f_{\rm ext}=-3\cdot 10^{-9} {\rm m/s}$. The  values of the other parameters are taken from Table \ref{tab1}.}
\label{pross}
\end{figure}

We have 
 plotted steady state profiles for the cell velocity, cell turnover, cell stress, and fluid pressure,  in 
 Fig. (\ref{pross}). 
 The fact that the cell velocity decreases continuously to zero as the ordinate goes to $h$, shows that cells die everywhere in the bulk. They divide only at the surface. This fact is also clear from the negative value of the turnover rate.
The high negative cell stress corresponds to a high pressure which is responsible for the large apoptosis rate. 

\begin{figure}[htb]
\includegraphics[width=5.2cm,height=4.5cm]{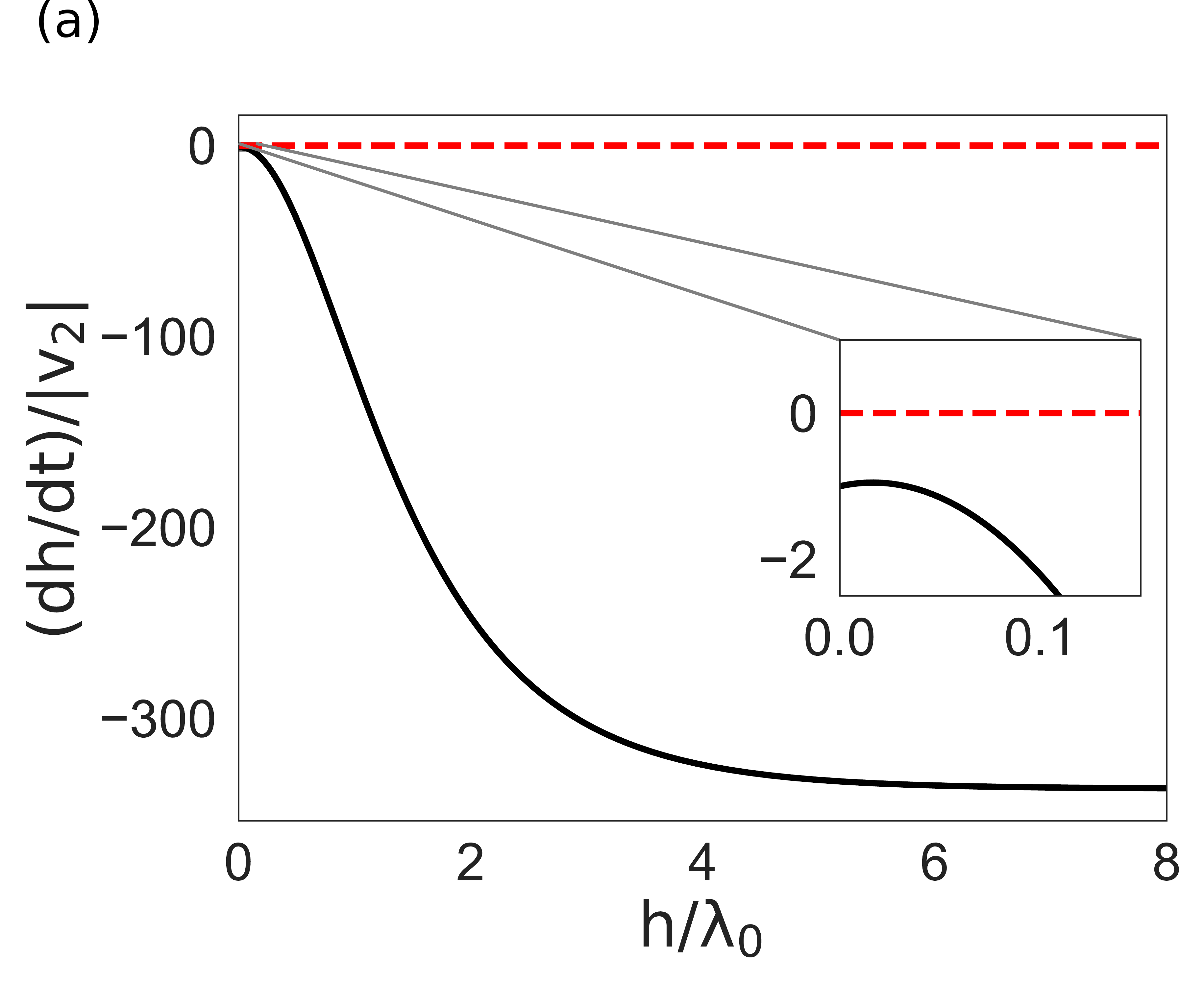}
\includegraphics[width=5.2cm,height=4.5cm]{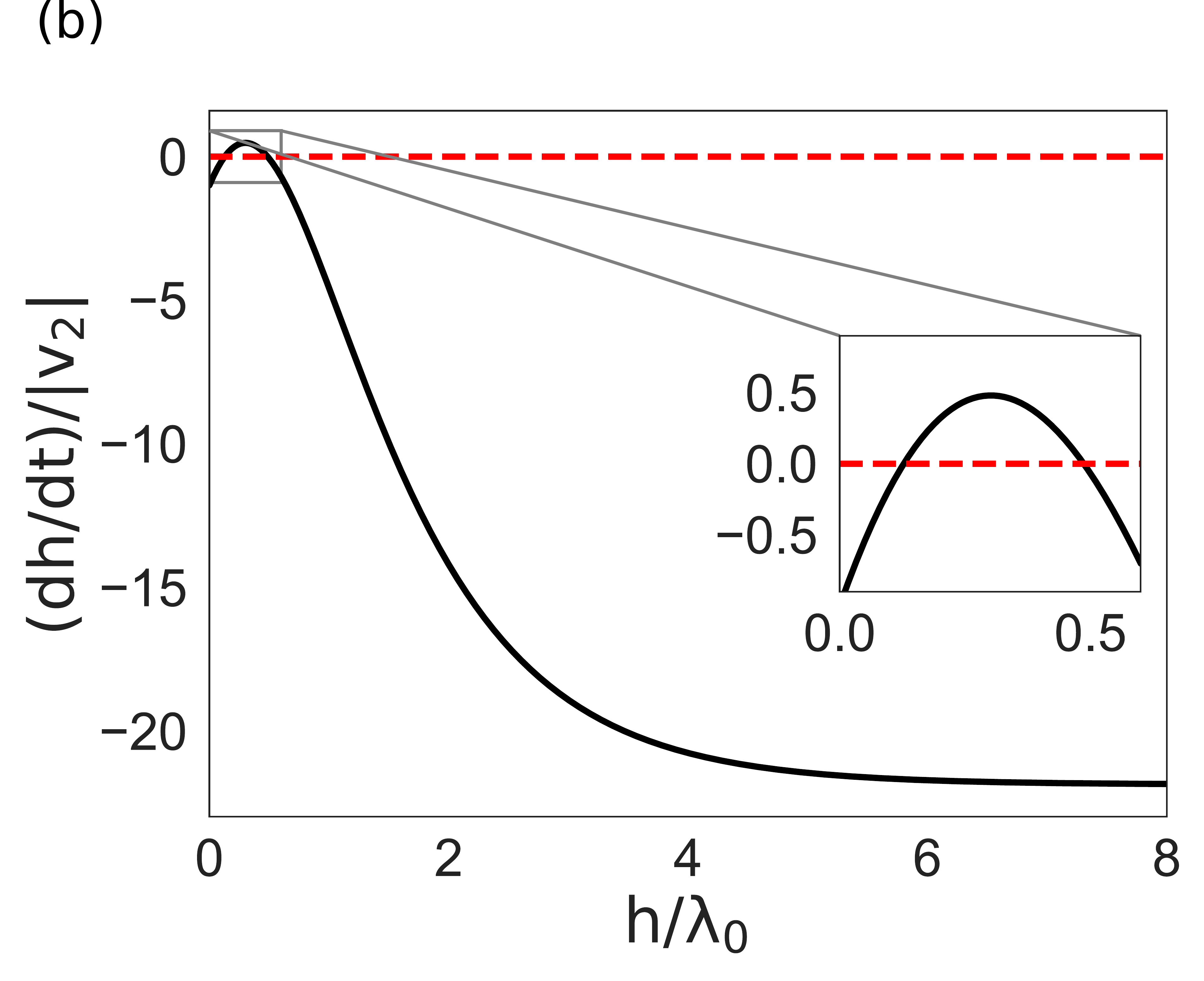}
\includegraphics[width=5.2cm,height=4.5cm]{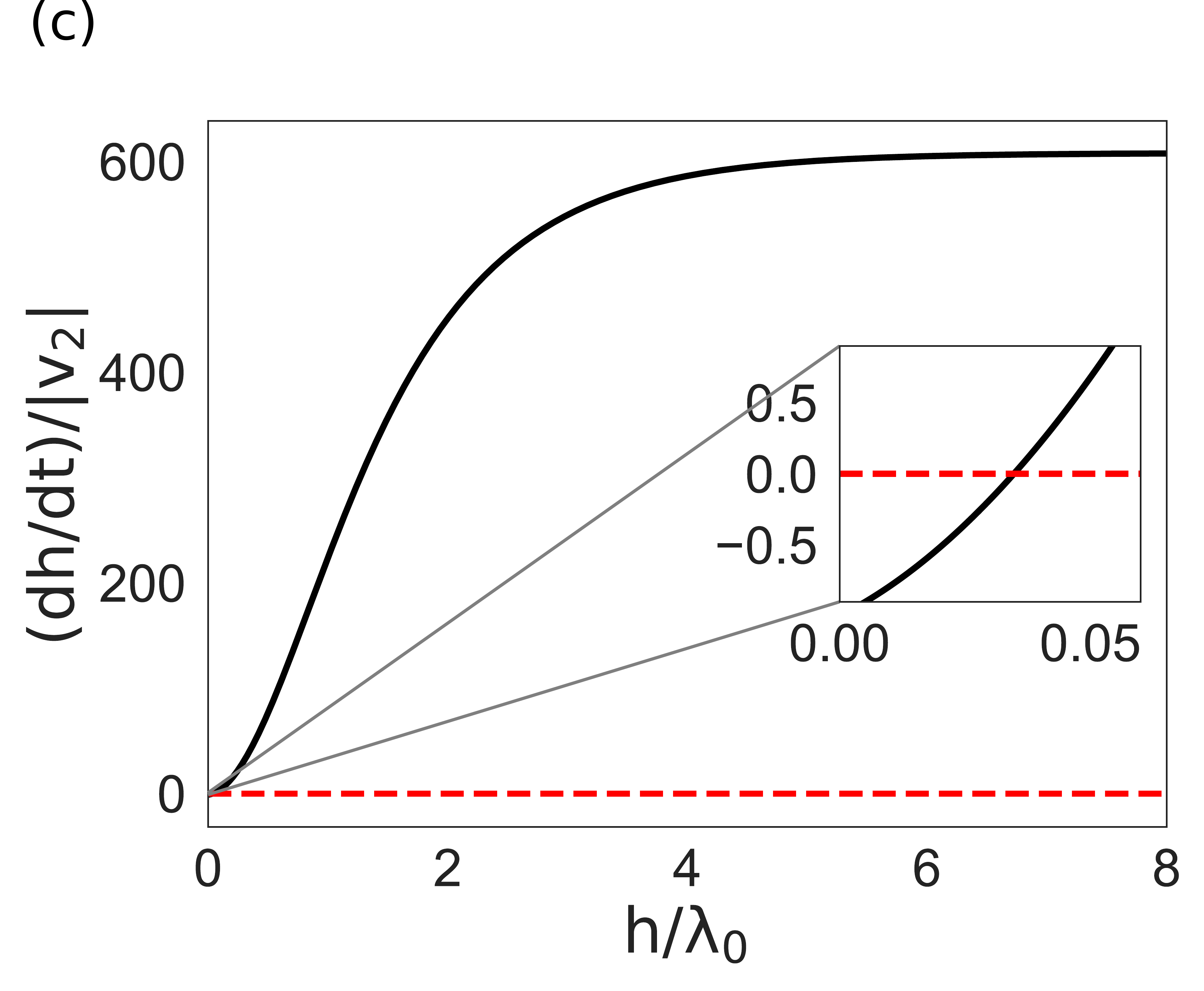}
\caption{
Plots of ${dh/ddt}$ vs ${h}$ for $v_{2}<0$ ($v_2=-3 \times 10^{-10} {\rm m/s}$) in the presence of electric current and fluid flow. (a) the layer collapses for a low electric current 
$I_{\mathrm{ext}}=2 {\rm A/m^2}$, (b) for a higher value of current $I_{\mathrm{ext}}=3 {\rm A/m^2}$, the tissue 
collapses for a small initial thickness whereas it reaches a steady state, with a large initial thickness, and, (c)  for higher value of 
current $I_{\mathrm{ext}}=5 {\rm A/m^2}$ it either collapses or proliferates depending on initial thickness. Here we have used $v^f_\mathrm{ext}=9 \times 10^{-8} {\rm m/s}$, and 
$P^c_h=-5 {\rm KPa}$. The values of the remaining parameters are taken from Table \ref{tab1}.}
\label{dhdtphase2}
\end{figure}

When $v_2<0$, the surface layer attached to the substrate is 
 a cell sink. 
  If the homeostatic pressure is also negative, the cells tend to die in the bulk too. What can prevent the 
tissue from an immediate collapse is 
 the opposite action of an electric field or of an external fluid flow 
 provided it has the right sign. We find 
that in the $dh/dt$ vs $h$ plot of Fig. (\ref{dhdtphase2}) for $v_2<0$, three different situations can 
arise, depending on the value of $I_{\mathrm{ext}}$ and $v^f_\mathrm{ext}$: (a) the tissue always
collapses with $dh/dt<0$, (b) with a pair of unstable-stable fixed points the tissue collapses for initial thicknesses smaller than  that of the unstable fixed point, and reaches a stable thickness for initial thickness values larger than the unstable one, (c) with one unstable fixed point, the tissue collapses for an initial thickness smaller than the thickness value of the unstable fixed point, and proliferates for an initial  thickness larger than that value. In Fig. 
(\ref{dhdtphase2}a) we see a total tissue collapse for a low value of electric current $I_{\mathrm{ext}}
=2 A/m^2$, which is indicative of the fact that the current and external fluid velocity are not enough to 
counter the large apoptosis rate due to the combined effect of negative homeostatic pressure and  surface apoptosis $v_2<0$. If the current is 
increased, keeping all other parameters constant, we find parameter values for which 
 the electric field 
   allows to obtain a stable tissue thickness provided the initial thickness is large enough. For a 
smaller initial thickness the tissue collapses as shown in Fig. (\ref{dhdtphase2}b).
If the current is increased further, 
tissue growth dominates, if the initial thickness is larger than that of the unstable fixed point 
and drives the tissue to an uncontrolled growth phase. For a smaller initial thickness, we still obtain a tissue collapse. 
This unstable steady state is shown in Fig. (\ref{dhdtphase2}c).

\begin{figure}[htb]
\includegraphics[width=16cm,height=10cm]{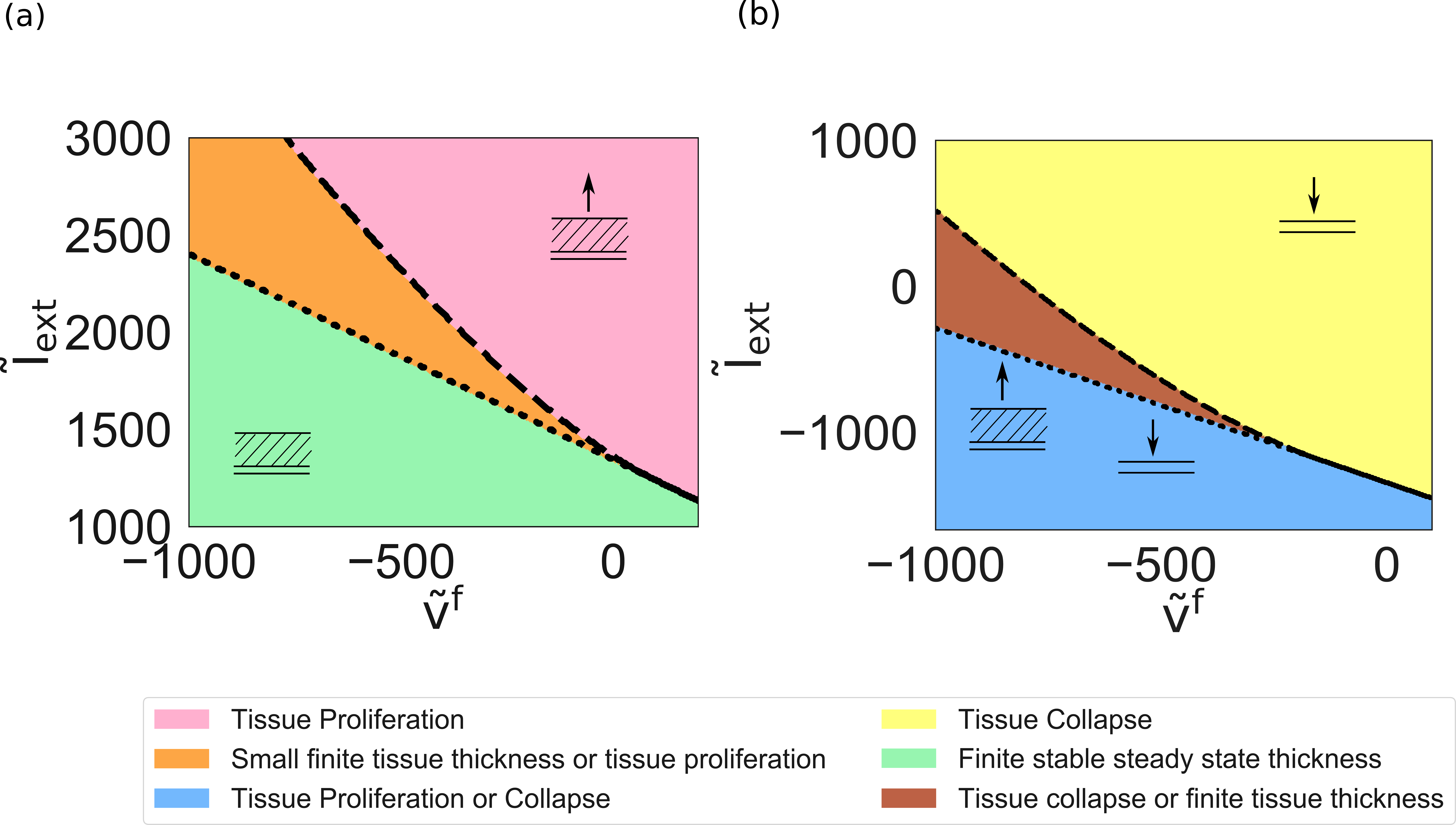}
\caption{State diagram of a thick epithelium in the presence of an external electric current and 
external fluid flow. 
(a) for $v_2=3 \times 10^{-10} {\rm m/sec}$, in the green domain the layer thickness evolves spontaneously  toward a stable finite value, in the pink domain the layer proliferate and in the orange domain they either evolve toward a finite stable value or proliferate depending on initial conditions; (b) for 
$v_2=-3 \times 10^{-10} 
{\rm m/sec}$ in the yellow domain the layer collapses, in the brown domain it either collapses or evolves toward a finite stable thickness and in the blue region it either proliferates or collapses depending on initial conditions. Here 
$\tilde v^f=v^f_\mathrm{ext}/v_2$, and 
$\tilde I_{\rm ext}=\lambda I_{\mathrm{ext}} (1-\phi)/(\Lambda \kappa_{\mathrm{eff}}v_2)$. 
We have used $P^c_h=-5 {\rm KPa}$. All other parameter values are taken from Table \ref{tab1}.} 
\label{phasefull1}
\end{figure}

These 
dynamical states for both positive and negative $v_2$ can be visualized in two seperate 
 diagrams 
in the $\tilde v^f$ - $\tilde I_{\rm ext}$ parameter space. 
We plot these 
 diagrams in Figs. (\ref{phasefull1}a) and (\ref{phasefull1}b) respectively. With $v_2$ 
kept constant, tuning $\tilde v^f$ 
can be achieved by tuning the external fluid velocity $v^f_\mathrm{ext}$, and 
tuning $\tilde I_{\rm ext}$ 
can be achieved by tuning the current $I_{\mathrm{ext}}$ flowing through the tissue, 
with all other parameters kept fixed.

Fig. (\ref{phasefull1}a) shows that for positive $v_2$, we find three 
scenarios: 
(a) green region: the tissue reaches a stable thickness irrespective of the initial conditions 
(b) orange region, existence of a pair of stable-unstable fixed points: for initial thickness values smaller than that of the unstable fixed point, the tissue slab goes to a stable finite thickness whereas for larger initial values the tissue proliferates 
(c) uncontrolled tissue proliferation ($h_{ss}\rightarrow\infty$) represented by the pink region. The solid line indicates a continuous transition between the green and the pink regions: approaching the line from the green side, the steady state thickness increases and diverges on the line. The dotted line between the green and the orange region signals the disappearance of the unstable fixed point of the orange region upon approaching the green region, the corresponding thickness diverging on the line. This is a spinodal line. The dashed line signals the simultaneous disappearance of the fixed point pair of the orange domain upon entering the pink domain. This is a line of discontinuous transition. The point where the three lines meet is again a tricritical point.

Fig. (\ref{phasefull1}b) 
illustrates the possible scenarios for negative $v_2$. We again find three 
possibilities in the $\tilde v^f$ - $\tilde I_{\rm ext}$ space: (a) the blue region corresponds to an unstable steady state, for initial thicknesses smaller than the fixed point value, the tissue collapses whereas for larger initial thicknesses it proliferates, 
(b) the brown region is defined by the existence of a pair of unstable-stable fixed points, for initial thickness values smaller than that of the unstable fixed point the tissue collapses and for larger initial values it goes to a stable finite thickness corresponding to the stable fixed point  (c) in the yellow region, $dh/dt$ is always negative and the tissue collapses. The solid line separating the blue and violet domains is a line of continuous transition: the unstable thickness increases continuously to infinity as  the tissue approaches the line from the blue side. The dotted line signals the disappearance of the stable fixed point as the corresponding thickness goes to infinity when the tissue enters the blue region coming from the brown side. This is a spinodal line. The dashed line signals a discontinuous transition as the pair of fixed points disappears when the tissue enters the yellow region coming from the brown side. 

\begin{table}[h!]
\begin{center}
\begin{tabular}{ |p{5cm}|p{5cm}| }
 \hline
 \multicolumn{2}{|c|}{Parameter list} \\
 \hline
 Parameter & Values  \\
 \hline
 $\eta$   & $10^4 \mathrm{Pa\cdot s}$    \\
 $\bar\eta$ &   $2\times 10^9 \mathrm{Pa\cdot s}$    \\
 $P^c_h$ & $
 -5000 \mathrm{Pa}$  \\
 $\phi$ &   $0.99$    \\
 $\kappa$    & $10^{13} \mathrm{Pa\cdot  s/m^2}$ \\
 $\bar\kappa$    & $10^{3} \mathrm{A\cdot s/m^3}$ \\
 $\zeta$ &   $
 -1000 \mathrm{Pa}$  \\
 $v_2$ & $
 20 \mathrm{\mu m/day}$  \\ 
 $\lambda_1$ & $-10^8 \mathrm{N/m^3}$ \\
 $\lambda$ & $10^5 \mathrm{N m^{-2}V^{-1}}$  \\
 $\Lambda_1$ & $3 \mathrm{A/m^2}$  \\
 $\Lambda$ & $10^{-3} \mathrm{S/m}$ \\
 $\Lambda^f$ & $10^{-11} \mathrm{ms^{-1}Pa^{-1}}$  \\
 $J_{\rm p}$ & $10^{-10} \mathrm{m/s}$ \\
 $\nu$ & $1$  \\
 $\nu_1$ & $1 \mathrm{Pa\cdot m/V}$  \\
 $\nu_2$ & $1.5 \mathrm{Pa\cdot m/V}$  \\
 \hline
\end{tabular}
\caption{Table of parameter values used in the plots.}
 \label{tab1}
\end{center}
\end{table}

\section{Discussion}

In this work, we have analysed the long term growth behaviour of planar thick epithelia permeated either by a constant fluid flow or by a constant electric current, or both. The predictions are striking, since one finds that the domain of stability of a finite thickness epithelium is rather small, and that  a simple dc electric current or a simple fluid flow is sufficient to either lead to tissue proliferation or to tissue collapse, without any need for genetic mutation. The results should be rather robust, since the above developed arguments are based on symmetry considerations, force conservation laws and cell number balance equations. 
Yet, this exercice would be futile if the field values required for observing these behaviours were out of experimental reach. Even though our phenomenological theory involves a rather large number of parameters, the formulation can be cast in a such a way as to involve only two control parameters, which can be estimated either from values already known experimentally, or from educated guesses. The one feature which is the most difficult to assess is the sign of the coupling parameters. These will need specific experiments to be pinned down. We expect clearly observable effects for flow fields or electric fields somewhat larger than those generated naturally in epithelia, but not orders of magnitude larger. The reason is that either flow or currents comparable to the naturally occuring ones can redistribute proteins in the cells and modify their polarity rather efficiently. The simplest result we obtain is that the steady state thickness of epithelia is proportional to the hydrodynamic screening length introduced in \cite{ranft2012} and which one can estimate from \cite{Delarue2014}, with a multiplicative logarithmic correction. The product turns out to be in the millimeter range, a very reasonable feature. The main limitation of this theory stems from the assumption of homogeneity in the direction parallel to the tissue layer. It is well possible that instabilities leading to lateral structuration of the tissue exist. This possibility should be investigated in the future. In any case, well controlled experiments are clearly needed, and would give us a deeper insight on the fundamental properties of thick epithelia. The prediction of either collapse or proliferation under suitable conditions is an exciting possibility which should be tested. Last, the process of tissue collapse may take an interesting twist: our analysis does not include explicitly the cell surface layer. It is only included in a flux boundary condition. As a result, the collapse may correspond to a thickness decrease up to the last layer, which usually has very different properties \cite{Delarue2014,monnier2016,montel2012}, or in other word to a transition from a thick epithelium to a monolayer epithelium. This aspect could be tested experimentally as well.

\section{Acknowledgements}

We thank Charlie Duclut for reading the manuscript and critical comments. NS would also like to thank Marko Popovic and Keisuke Ishihara for fruitful discussions.

\section{Appendix}

\subsection{Estimation of relevant parameters}

One can obtain an order of magnitude of the coefficient $\lambda=\lambda_2+2\lambda_3/3$, by estimating the 
shear stress created by the electo-osmotic flow in the intercellular region. 

The hydrodynamic stress is as usual given by $\sigma=\eta \partial_x v$, where $\eta$ is the interstitial 
fluid viscosity, and $\partial_x v$ the shear rate due to electro-osmosis at the cell membrane surface. 
Standard calculations \cite{kirby2010} yield
\bea
\partial_x v \simeq {\epsilon \epsilon_0 \zeta_E \over \lambda_D\eta} E_z^{\mathrm{cleft}},
\eea
where $\epsilon$ is the permittivity of the fluid (here water), $\epsilon_0$ is the permittivity of vacuum, 
$\zeta_E$ is the zeta potential of the cell membrane, 
$E_z^{\mathrm{cleft}}$ is the electric field 
in the cleft, and $\lambda_D$ is the Debye screening 
length. 

\begin{figure}[htb]
\includegraphics[height=8cm]{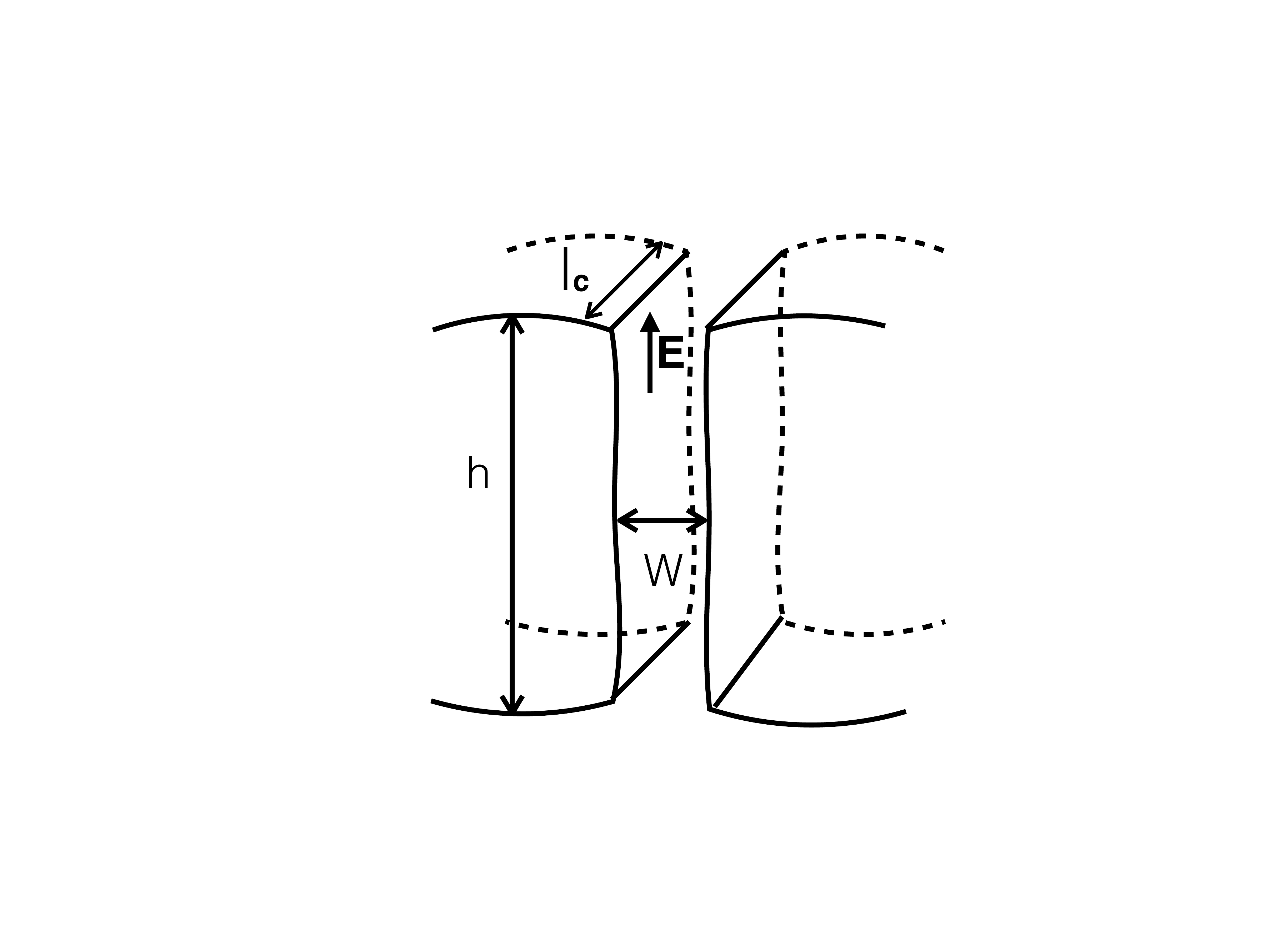}
\caption{Schematic diagram showing a curved tissue and interstitial space between the cells}
\label{ld}
\end{figure}

Then the stress is expressed in terms of $E_z^{\mathrm{cleft}}$ as
\bea
\sigma\simeq {\epsilon \epsilon_0 \zeta_E \over \lambda_D} E_z^{\mathrm{cleft}}, \label{cleft}
\eea 
which gives the total force on the cell to be $4 l l_c\sigma$, with $l$ being the height of the cell, and 
$l_c$ being the width of the cell. The force per unit volume can then be written as
\bea
f_E\simeq{4 l l_c\sigma \over l l_c^2}= {4\epsilon \epsilon_0 \zeta_E \over \lambda_Dl_c} E_z^{\mathrm{cleft}}.
\eea

The relation between the coarse grained electric field $E_z$, and the actual field $E_z^{\mathrm{cleft}}$ 
in the intercellular domain is simply given by the geometrical relation $E_z^{\mathrm{cleft}}\simeq l_cE_z/(2W)$ stemming from current conservation ignoring the current passing through the cells, 
where $W$ is the width of the intercellular cleft. Using this relation, the force per unit volume 
can be
expressed as
\bea
f_E\simeq {2\epsilon \epsilon_0 \zeta_E \over \lambda_DW} E_z,
\eea
which leads to
\bea
\lambda\simeq{2 \over W}{\epsilon \epsilon_0 \zeta_E \over \lambda_D}. 
\eea 
Using $\epsilon_0=8.85\cdot 10^{-12} \mathrm{F/m}$, $\epsilon\simeq 80$ for water, 
$\zeta_E\simeq 3\cdot 10^{-2} \mathrm{mV}$, 
$W\simeq 10^{-7} \mathrm{m}$, and $\lambda_D\simeq 10^{-9} \mathrm{m}$, we obtain 
$\lambda\simeq 3 \cdot 10^{5} 
\mathrm{N m^{-2}V^{-1}}$.

The term $\lambda_1$ in (\ref{falpha}) corresponding to cell polarity has also never been measured to our knowledge. 
If we assume that it is comparable to the electric term, under conditions such that the average flux 
vanishes, we can infer an estimate based on the knowledge of typical potential differences generated by 
epithelial cells i.e., mV. Then $\lambda_1\simeq - \lambda E_z$. As the potential difference 
$V\simeq 10^{-3} V$ in MDCK blisters \cite{misfeldt1976,cereijido1978}
 and width of a cell $l_c\simeq 10^{-5} \mathrm{m}$, then $E_z \simeq V/l_c \simeq 
\cdot 10^2 \mathrm{V/m}$. Thus $\lambda E_z\simeq - \lambda E_z \simeq -3 \cdot 10^{5}10^{-3}/10^{-5} 
\mathrm{N/m^3} \simeq -10^8 \mathrm{N/m^3}$

We next estimate the values of $\nu$, $\nu_1$, and $\nu_2$ in equations (\ref{isostressform}) and 
(\ref{anisocellstressform}). We argue that $\nu$ should be of $O(1)$, as it is a dimensionless parameter as 
evident from (\ref{isostressform}). To determine the value of $\nu_1$, we assume that $\nu_1p_\alpha E_z$ 
plays the role of a pressure, even though the underlying physics might be more subtle. Thus we equate 
$\nu_1p_\alpha E_z= \lambda h E_z$, so that $\nu_1=\lambda h\sim O(1) \,\,\mathrm{Pa\cdot m/V}$. For 
$\nu_2$, we compare Eq. (\ref{anisocellstressform}) with (\ref{cleft}), from which we can express 
$\nu_2\simeq \epsilon \epsilon_0\zeta_El_c/(2W\lambda_D) =80\cdot 8.85\cdot 10^{-11}\cdot 3\cdot 10^{-2}/10^{-9}\cdot 50 
\mathrm{Pa\cdot m/V} \simeq 1.5 \mathrm{Pa\cdot m/V}$.


The coefficient $\bar\kappa$ may be estimated using standard  streaming potential relations \cite{hubbard2002}  adapted 
to the intercellular domains, similar in spirit to what was done for estimating $\lambda$. We obtain
\bea
\bar\kappa \simeq {12 \epsilon \zeta_E  \over W l_c}. \label{kappabar}
\eea
Using the values above and $l_c\simeq 10^{-5} \mathrm{m}$, gives $\bar\kappa\simeq 10^{3} 
\mathrm{A\cdot s/m^3}$.

The coefficient $\Lambda^f$ can also be estimated using standard fluid flux calculated in the intercellular 
cleft and from geometric relations, just like we used in the calculation of $\lambda$, we obtain the 
expression 
\bea
\Lambda^f\simeq {W^3 \over 6 l_c^2\eta}.
\eea 
Using the values of $W$, $l_c$, and $\eta$, given above, yields $\Lambda^f\simeq 10^{-11} \mathrm{m/(Pa\cdot 
s)}$.

To estimate the value of $J_p$ we note that at steady state the pressure difference $P^1_{\rm ext}-P^f$ 
is equal to the osmotic pressure difference $\Pi_{\rm ext,0}^1-\Pi_{\rm int,0}^1$, where 
$J_p=\Lambda^f(\Pi_{\rm ext,0}^1-\Pi_{\rm int,0}^1)$. Furthermore we know that the curvature radius of a cell membrane $R$ 
should be much larger than the cell thickness $l_c$, for the cell to be stable. So the pressure difference 
$P^1_{\rm ext}-P^f\simeq \gamma/R$ should be much smaller than $\gamma/l_c$. This ensures, 
$\Pi_{\rm ext,0}^1-\Pi_{\rm int,0}^1\leq \gamma/l_c$. Using the values of $\gamma\simeq 10^{-4} \mathrm{N/m}$, 
and $l_c\simeq 10^{-5} \mathrm{m}$, we obtain $J_p\simeq 10^{-10} \mathrm{m/s}$.


\end{document}